\documentclass[twocolumn,preprintnumbers,amsmath,amssymb,aps,pre,reprint,longbibliography]{revtex4-2}
\usepackage{graphicx}
\usepackage{xcolor}
\begin{document}

\title{Hysteresis, Laning, and Negative Drag in Binary Systems with Opposite and Perpendicular Driving}
\author{
C. Reichhardt and C. J. O. Reichhardt 
} 
\affiliation{
Theoretical Division and Center for Nonlinear Studies,
Los Alamos National Laboratory, Los Alamos, New Mexico 87545, USA
}

\date{\today}

\begin{abstract}
  We consider a binary system of particles with repulsive interactions that move in opposite or perpendicular directions to each other under an applied external drive. For opposite driving, at higher drives a phase-separated laned state forms that has strong hysteresis in the velocity-force curve and the fraction of topological defects as the drive is cycled up and down from zero. The amount of hysteresis depends on the drive value at which the drive changes from increasing to decreasing. For perpendicular driving, we find a jammed state that transitions into a disordered state or a tilted lane state, both of which also show strong hysteresis effects. Additionally, a negative drag effect can appear in which one species moves in the direction opposite to the other species due to a tilting of the lanes by the perpendicular drive. When a constant drive is applied along one direction while the drive in the perpendicular direction is increased, we observe a series of drops and jumps in the velocity as the system forms locked and tilted laned states. For weakly interacting particles, the jammed system can show co-tilted stripe-forming states.
\end{abstract}

\maketitle

\section{Introduction}

Numerous studies have addressed
systems containing two species of repulsive particles that
move in opposite directions under an applied drive.
This includes studies of
Yukawa particles \cite{Dzubiella02a,Chakrabarti04,Glanz12},
particles with short-range repulsion \cite{Ikeda12,Klymko16,Poncet17,Yu24},
binary colloidal systems
\cite{Leunissen05,Rex07,Lowen10,Vissers11a,Geigenfeind20},
colloids with different mobility in gravity fields \cite{Isele23},
models of pedestrian flow 
or social systems \cite{Helbing00,Feliciani16,Sieben17,Bacik23},
dusty plasmas \cite{Sutterlin09,Du12},
hard disk systems \cite{Reichhardt18},
binary active matter \cite{Bain17,Reichhardt18b,Khelfa22},
and skyrmion-skyrmionium mixtures \cite{Vizarim25}. 
One of the most interesting effects that arises in these systems
is phase separation into well-defined lanes of oppositely moving particles.
Typically, at low drives, the particles are
in a low velocity chaotic mixed state, while at high drives, rapidly
moving phase separated lanes appear. The amount of ordering in the system
increases once the lanes form,
and there can be a jump up in the absolute velocity due to a reduction
in collisions between oppositely moving particles
\cite{Reichhardt18}.
Several dynamical phases appear,
including a jammed phase where the particles cannot move past each other,
disordered or chaotic flow, mixed laning,
and ordered laning
\cite{Dzubiella02a,Reichhardt18,Glanz12,Wachtler16,Yu22}. 

Laning transitions can also occur for binary mixtures
under ac driving \cite{Wysocki09,Li21} and when
the particle-particle interaction rules are modified.
For the case of oppositely moving pedestrians, when a rule was introduced
that caused particles to move preferentially to one side when interacting
with an oppositely moving particle,
a chiral effect emerged in which
particle separation into tilted lanes occurred
\cite{Bacik23}. 
In studies of oppositely driven systems where a Magnus force is present
that generates a velocity component perpendicular to the net force
experienced by a particle, similar chiral effects lead to
the formation of tilted lanes
as well as different types of flow phases \cite{Reichhardt19a}.
In skyrmion-skyrmionium mixtures,
where skyrmions experience a Magnus force and skyrmioniums do not,
the two species move at different velocities and
different angles with respect to the applied drive, again causing
the formation of tilted particle lanes \cite{Vizarim25}.
For systems where only a portion of the particles are driven while the
remaining particles are mobile but undriven,
an effective attraction
can arise
between the driven and undriven mobile particles
\cite{Dzubiella03,Reichhardt06,MejiaMonasterio11,Poncet19}. 

Studies of laning in oppositely driven particles
have generally focused only on the type of ordering that appears
under an increasing drive, but
an open question is whether hysteresis can occur
if the drive is then reduced back to zero.
For example, if the initial non-driven state is uniformly mixed
and the high-drive state is laned, does the laned state
persist to lower drives as the drive is reduced back to zero,
and would hysteresis of this type produce signatures
in the transport curves for the two species?
For driven systems that undergo nonequilibrium transitions from a disordered
state to a moving smectic or moving crystal
state at high drive,
such as vortices in type-II superconductors  \cite{Reichhardt97}
and other particle assemblies \cite{Reichhardt17},
it is possible for hysteresis to occur in which the ordered state persists
down to drives below the drive at which ordering occurred during the
upward drive sweep.
In general, however,
for driven particle systems moving over
random quenched disorder in two dimensions,
hysteresis in the dynamic reordering
is not observed \cite{Olson98a,Reichhardt17}.
Hysteresis
in the laning transition for oppositely driven particles 
would not be expected upon cycling the drive up and down if the transition is
second-order or a crossover, but would occur
if the laning transition is first-order in character.
For the case of
a binary assembly of repulsive particles that are identical except for the
direction in which each particle is driven,
the ground state in the absence of driving is a triangular lattice,
and phase separation into lanes occurs only in the driven state.

Beyond the well-studied dynamic reordering into
crystalline states, relatively little
is known about what types of
dynamic phases or laning states
could occur for more exotic
driving protocols,
such as if one particle species is driven in a direction perpendicular, rather
than opposite to, the driving direction of the other species.
This could be achieved by driving one species in
the positive $x$-direction and the other species in the positive
$y$-direction. A situation of this type can arise
in certain types of pedestrian flows,
or for cases in which one particle species
is driven by gravity and the other
is driven by an electric field.
Studies of pedestrian motion
under perpendicular or crossed flows have shown the formation of
jammed states and tilted lane states \cite{Cividini13}.
There have also been studies of three-dimensional
dusty plasmas with perpendicular driving that have demonstrated the
formation of tilted laned states \cite{Prajapati23}.
It would be interesting to
determine whether laning systems with
perpendicular drives show hysteresis
and whether the type of laning
that occurs is different from that found
for oppositely driven particles.

In this work, we consider the effects of changing the relative orientation
of the driving direction in a
binary assembly of repulsive particles.
The particles are driven either in opposite or in perpendicular directions,
and we sweep the drive up from zero to a maximum value before sweeping
the drive back down to zero.
When the particles are oppositely driven,
we find a jammed state, a fluctuating state, and
a disordered state as a function of increasing drive.
When the drive becomes large
enough for the system to form a laned state,
we observe strong hysteresis in both the velocity-force curves
and the fraction of particles with six-fold ordering.
In this case, the laned state persists to a much lower drive during
the ramp-down of the drive, and the system exhibits much more
triangular ordering than it had during the fluctuating state on
the upward ramp of the drive.
Even when the drive has been brought back down to zero,
the ground state energy of the previously laned state
is the same as that of a mixed state with triangular ordering,
but lower than that of a topologically disordered state.

For perpendicular driving, we find a locked or jammed phase at
low drives in which the particles attempt to move along the
$45^\circ$ direction, followed by a decoupling transition at which
disordered flow occurs.
At higher drives, various tilted lane states appear with transitions
that produce upward and downward jumps in
the transport curves.
There is strong hysteresis,
and the tilted lane state persists to much lower drives
as the drive is swept back to zero.
We also find a negative drag effect
that occurs for both species of particles,
and that arises when one of the particle species
travels in the direction opposite
to the driving direction of the other species, despite the fact that the
applied driving directions are perpendicular.
This negative drag effect occurs due to the tilting of the lanes,
and appears when the particles
are able to run along a lane that is
tilted in the opposite direction to the drive of the other species.
We also consider the effect of fixing the driving of one species along
the $x$ direction while slowly increasing the driving of the second
species in the perpendicular or $y$ direction.
In this case, there is a coupling-decoupling transition
accompanied by the formation of different
types of tilted lanes.
Depending on the tilt of the lanes,
the hysteresis can be positive, where the particle velocities are higher
on the ramp-down of the drive than on the ramp-up, or negative, where
the velocities are higher during the ramp-up.
The transitions between the different laned
states produce
jumps in the number of topological defects in the system.

\section{Simulation}

We consider a two-dimensional system of size $L \times L$ with $L=36$ and
with periodic boundary conditions in the $x$ and $y$ directions.
Within the sample, we place $N$ particles that
have a repulsive Coulomb pairwise interaction
potential of $V(R_{ij}) = Q/R_{ij}$,
where $R_{ij} = |{\bf R}_i - {\bf R}_j|$
is the distance between particles $i$ and $j$ and $Q$ is the
force prefactor.
In this work we fix $Q=1$.
The system is evenly divided with $N/2$ particles of
species A, which are assigned $\sigma_i=1$,
and $N/2$ particles of species B, which are assigned $\sigma_i=0$.
The total particle density is $\rho = N/L^2$, and at zero temperature, this
system will form a triangular lattice.
We initialize the particle positions
by placing the particles in their
native triangular lattice.
After the initialization, we apply an external driving force of
magnitude
$F_D^A$ and direction $\bf{\hat a}=\bf{\hat x}$ for species A and,
for species B, a magnitude $F_D^B$ and either
$\bf{\hat b}=-\bf{\hat x}$ for opposite driving or
$\bf{\hat b}=\bf{\hat y}$ for perpendicular driving.
The overdamped equation of motion for the particles is:
\begin{equation}
  \eta \frac{d {\bf R}_{i}}{dt} = -\sum^{N}_{j \neq i} \nabla V(R_{ij}) +
       {F}_{D}^A \delta(\sigma_i-1) {\bf \hat a}
       +
       F_D^B \delta(\sigma_i) {\bf \hat b}.
\end{equation}
Here, $\eta$ is the damping coefficient, which we set to $1.0$.
For the first part of this work, we fix $F_D^A=F_D^B=F_D$.
We start with $F_D = 0.0$ and gradually increase $F_D$ in increments of $\Delta F_D = 0.001$ every $4 \times 10^3$ to $10^4$ molecular dynamics
simulation time steps. Once the drive has reached a maximum
value of $F_D^{\rm max}$, we begin decrementing the drive at the same
rate with $\Delta F_D=-0.001$ until the system has returned to $F_D=0.0$.
We also separately consider the case
in which the $x$ direction drive is held fixed at $F_D^A$ while
a perpendicular drive $F_D^B=F_D$ is gradually increased from zero.
We measure the time-averaged velocity for each species
in both the $x$ and $y$-directions:
$\langle V^A_x \rangle = (2/N)\sum^{N}_{i=1}\delta(\sigma_{i} -1)({\bf v}_{i}\cdot \hat{\bf x})$,
$\langle V^A_y \rangle = (2/N)\sum^{N}_{i=1}\delta(\sigma_{i} -1)({\bf v}_{i}\cdot \hat{\bf y})$,
$\langle V^B_x \rangle = (2/N)\sum^{N}_{i=1}\delta(\sigma_{i})({\bf v}_{i}\cdot \hat{\bf x})$,
and $\langle V^B_y \rangle = (2/N)\sum^{N}_{i=1}\delta(\sigma_{i})({\bf v}_{i}\cdot \hat{\bf y})$, where ${\bf v}_{i}$ is the velocity of particle $i$.
We also measure the fraction of sixfold-coordinated particles
$P_6=(1/N)\sum^{N}_{i=1}\delta(z_i-6)$, where $z_i$ is the coordination
number of particle $i$ obtained from a Voronoi tesselation in which all
particles are treated as identical regardless of species.

We treat the long-range repulsion with a Lekner summation method for 
computational efficiency \cite{Lekner91,GronbechJensen97a}.
In previous work, we
used the same model of Coulomb interacting particles in which all of the
particles were of a single species
to study depinning and sliding under dc driving when
the particles
were also interacting with a random substrate \cite{Reichhardt22}. 
A variation of this model with both
Coulomb repulsion and short-range attraction
was also used to study pattern formation and 
dynamics on periodic one-dimensional (1D) substrates \cite{Reichhardt25}.
In this work, we vary $\rho$, $F_D$, and $F_D^A$.

\section{Oppositely Driven Particles}	

\begin{figure}
\includegraphics[width=\columnwidth]{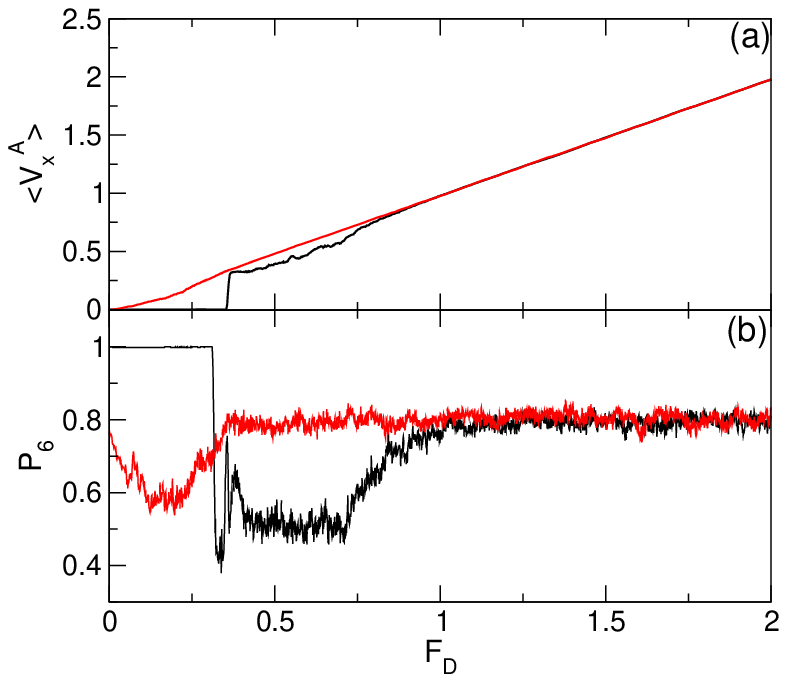}
\caption{
(a) The average $x$ velocity $\langle V^A_x \rangle$ vs
drive force $F_D$ for particles driven in opposite directions
at a density of $\rho = 0.441$.
The black curve represents the ramp-up phase,
and the red curve represents the ramp-down phase.
(b) The fraction of particles with six neighbors,
$P_6$, vs $F_D$ during ramp-up (black) and ramp-down (red).
We find three distinct states:
a jammed state, a disordered state, and a laned state at high
driving. There is strong hysteresis across the laning transition.
}
\label{fig:1}
\end{figure}

\begin{figure}
\includegraphics[width=\columnwidth]{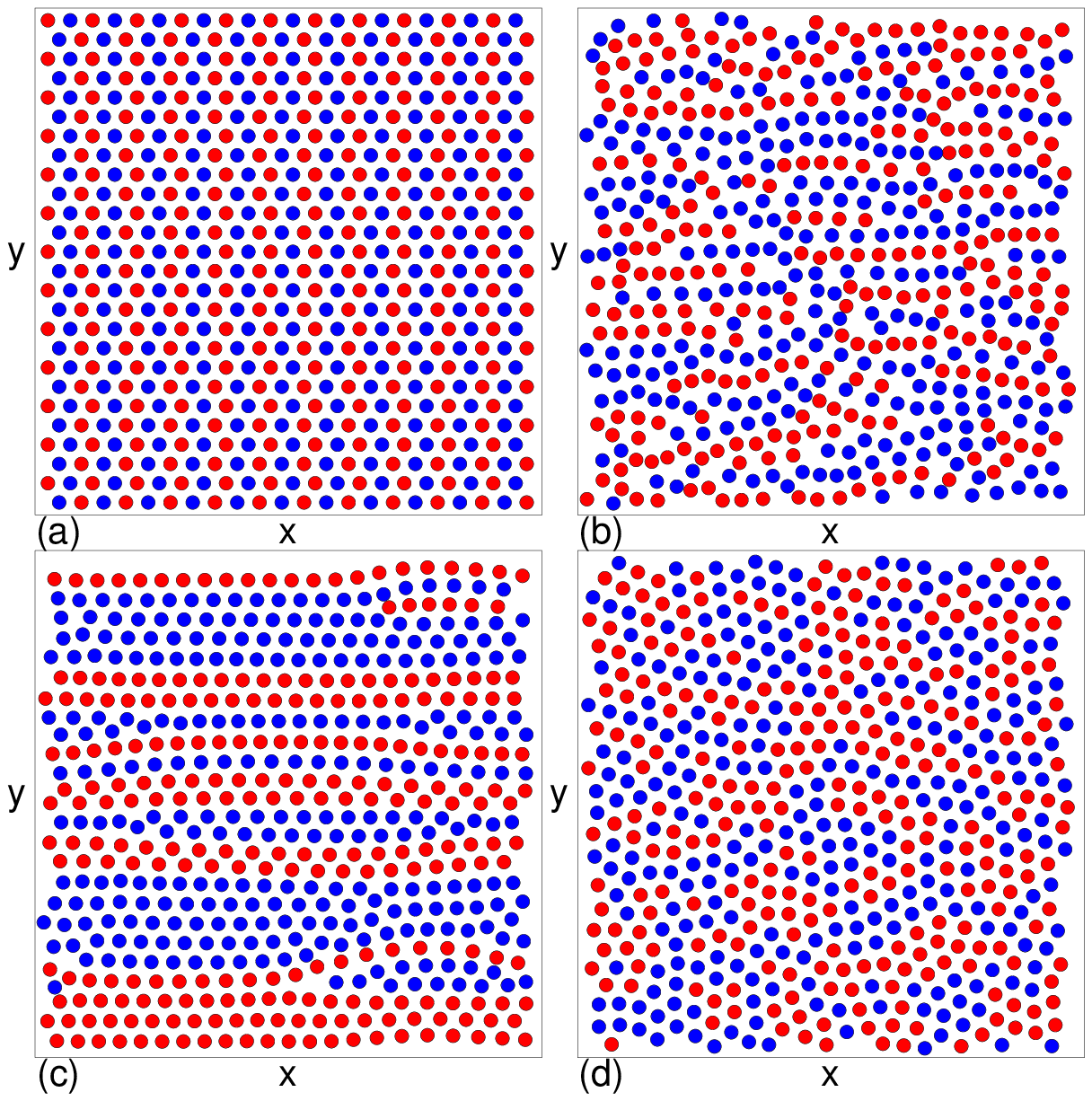}
\caption{Particle positions for the $\rho=0.441$ oppositely driven system
from Fig.~\ref{fig:1}. Species A (blue) is driven along $+x$, and
species B (red) is driven along $-x$.
(a) The initial jammed state at $F_D = 0.25$,
where the system forms a triangular lattice.
(b) The unjammed disordered state at $F_D = 0.5$,
where the triangular order is lost.
(c) The laned state at $F_D = 2.0$.
(d) The $F_D = 0.0$ jammed state after the ramp-down is completed.
}
\label{fig:2}
\end{figure}

We first consider hysteresis for oppositely driven particles in
a sample with a density of
$\rho = 0.441$.
In Fig.~\ref{fig:1}(a), we plot $\langle V_x^A \rangle$, the species A
velocity,
versus $F_D$ in black for the ramp-up and in red for the ramp-down portion
of the cycle.
Figure~\ref{fig:1}(b) shows 
the fraction of particles with six neighbors, $P_6$,
versus $F_D$ for all particles during the ramp-up and ramp-down.
On the ramp-up phase, the system is in a jammed state with
$\langle V_x^A \rangle = 0.0$
up to $F_D = 0.365$, and for these low drives, the two particle species are
unable to move past each other.
In Fig.~\ref{fig:2}(a), we illustrate
the particle configurations in the jammed phase at $F_D = 0.25$.
The system is initialized as
a triangular solid with $P_6 = 1.0$.
Once $F_D > 0.365$, the particles are able to
move past one another or unjam,
and for $0.365 < F_D < 1.0$,
we find a disordered or fluctuating phase.
Here the particles are moving around one another,
the lattice structure is lost,
and there is a generation of topological defects,
as shown in Fig.~\ref{fig:2}(b) at $F_D = 0.5$,
where the overall particle positions are disordered
and $P_6 \approx 0.5$.
For $F_D > 1.0$, the system organizes into an ordered laned state,
as shown in Fig.~\ref{fig:2}(c) at $F_D = 2.0$.
There is phase separation into elongated regions of
oppositely moving particles, and patches of sixfold ordering begin
to appear, causing $P_6$ to increase up to $P_6\approx 0.8$.
This laned state is not completely ordered,
since there are some dislocations in the
lattice near the separation between
oppositely moving lanes.
A similar laned state for the ramp-up phase
has been observed previously in
oppositely driven particle systems with repulsive interactions
\cite{Dzubiella02a,Chakrabarti04,Glanz12,Ikeda12,Klymko16,Poncet17,Reichhardt18,Yu24}.  

We find that the absolute velocity of any one particle is higher
in the laned state than in the fluctuating state,
since direct collisions between the particles are reduced
once the lanes form, leading to a smaller drag on the particles.
During the ramp-down of the drive,
the system remains in a laned state to a lower drive
of $F_D = 0.28$.
This is visible in the $P_6$ versus $F_D$ curves in Fig.~\ref{fig:1}(b),
where the amount of order remains much larger above the jamming threshold
during the ramp-down phase than it was in the
ramp-up phase.
Similarly,
in Fig.~\ref{fig:1}(a),
$\langle V_x^A \rangle$ at lower drives is
larger for the ramp-down phase than for the ramp-up phase.
During the ramp-down,
the system does not return to the same ordered, jammed state
illustrated in Fig.~\ref{fig:2}(a),
but remains in a fluctuating, fluid state
until it reaches the low drive, partially disordered jammed state,
shown in Fig.~\ref{fig:2}(d) at $F_D = 0.0$.
At the lowest values of $F_D$ during the ramp-down, the amount of
topological order present is greater than in the
fluctuating lattice state since the repulsive particle-particle interactions
favor the formation of
a triangular lattice, but full order is not restored since
some dislocations become trapped after the system passes
through the disordered phase during the ramp-down.
If we ramp-up the drive for a second time
after the ramp-down is complete,
we observe similar hysteresis in
the laned state.

\begin{figure}
\includegraphics[width=\columnwidth]{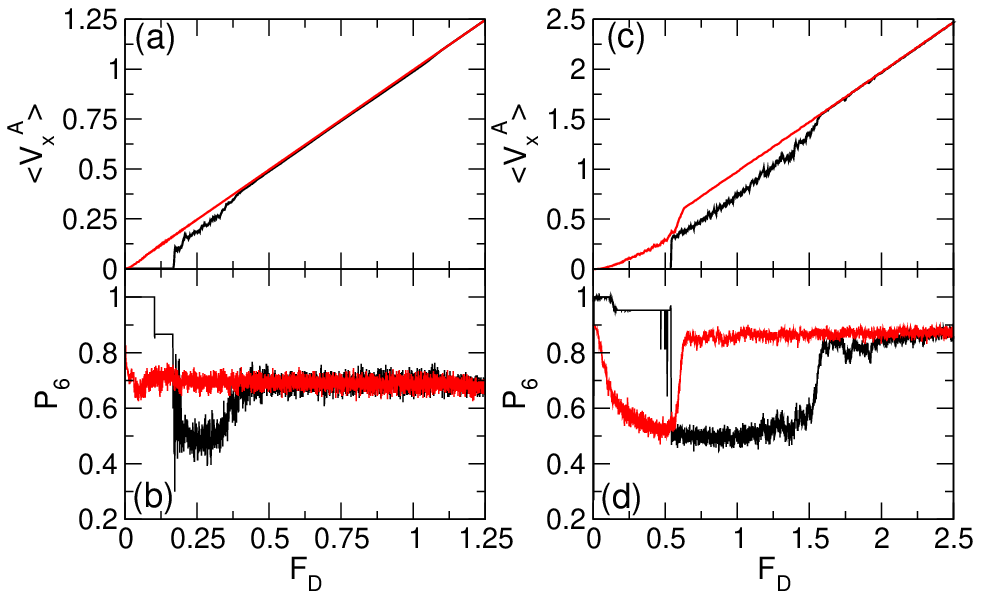}
\caption{
(a) $\langle V_x^A \rangle$ vs $F_D$ for
oppositely driven particles at $\rho = 0.208$.
(b) The corresponding $P_{6}$ vs $F_{D}$.
(c) $\langle V_x^A \rangle$ vs $F_{D}$ for
oppositely driven particles at $\rho = 0.93$.
(d) The corresponding $P_{6}$ vs $F_{D}$.
Black curves are for ramp-up and red curves are for ramp-down.
For both densities, the response is strongly hysteretic.
}
\label{fig:3}
\end{figure}

At other values of $\rho$, similar hysteresis effects appear.
In general, as $\rho$ increases,
the drive at which the laning state appears during the ramp-up shifts
to higher values of $F_D$.
In Fig.~\ref{fig:3}(a,b),
we plot $\langle V_x^A \rangle$ and $P_6$ versus
$F_D$ for an oppositely driven system
with $\rho = 0.208$,
and Fig.~\ref{fig:3}(c,d) shows
$\langle V_x^A \rangle$ and $P_6$ versus $F_D$ for oppositely
driven particles at $\rho = 0.93$.
When $\rho = 0.208$, the window of disordered flow is narrower,
and the system organizes to a laned state near $F_D = 0.5$
where $P_6$ reaches a value close to
$P_6=0.75$.
During the ramp-down,
the laned state persists all the way
to $F_D = 0.0$,
and there is almost no jammed phase present.
For $\rho = 0.93$, on the ramp-up
the jammed phase persists up to $F_D = 0.05$,
and the laned state forms at $F_D = 1.5$.
As the drive is ramped back down,
the laned state persists down to $F_D = 0.5$
before the system disorders.
There is a velocity drop at the onset of the
disordered phase,
but the velocity still remains higher during the ramp-down than
during the ramp-up since
the system does not lose all of the partial phase separation
it accrued during the formation of the laned state.

\begin{figure}
\includegraphics[width=\columnwidth]{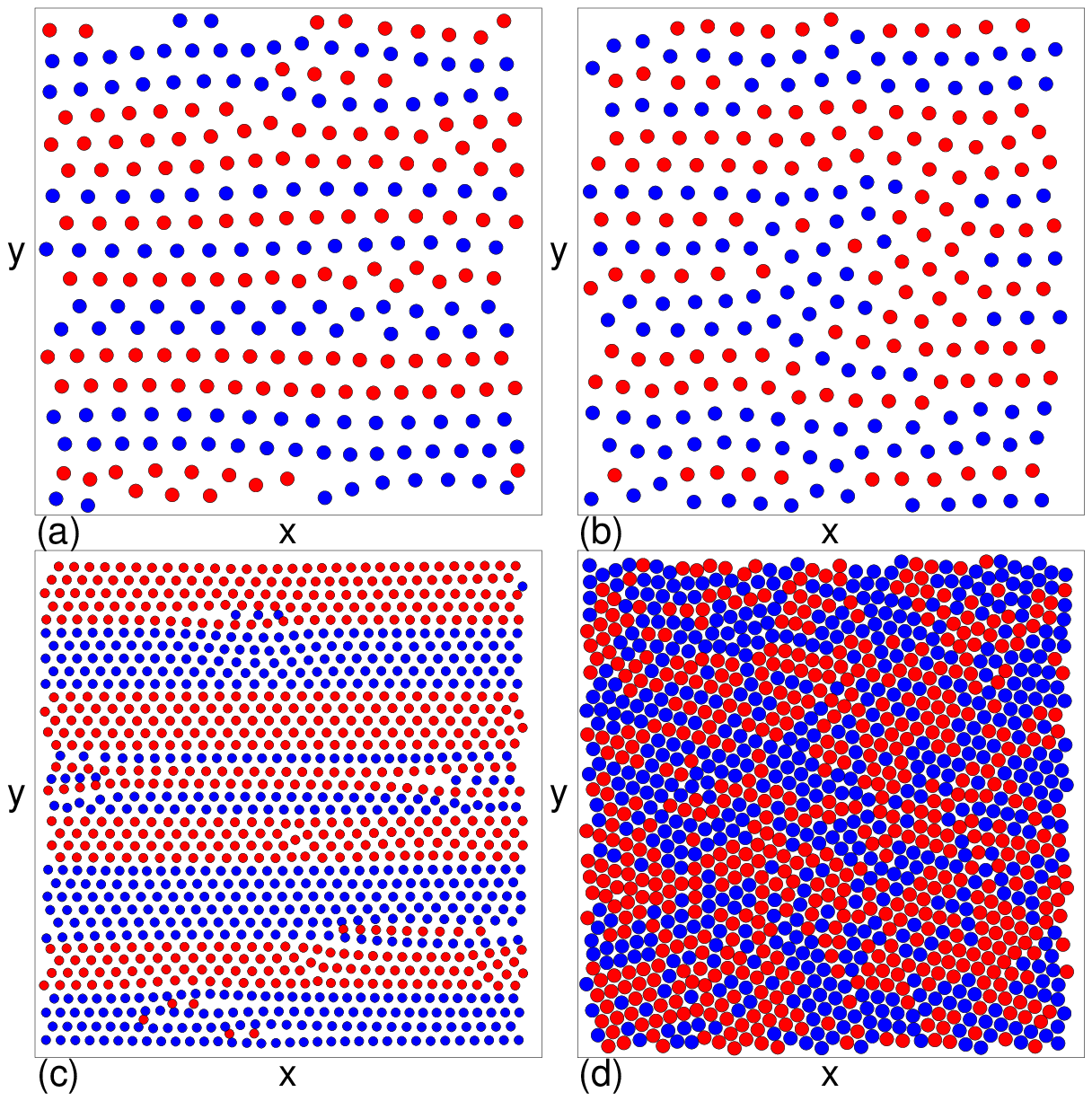}
\caption{Particle positions for an oppositely driven
system where species A (blue) is driven along $+x$ and
species B (red) is driven along $-x$.
(a,b) A system with $\rho=0.208$.
(a) The laned state at $F_D=1.0$.
(b) The jammed state at $F_D=0.0$ after the ramp-down.
(c,d) A system with $\rho=0.93$.
(c) The laned state at $F_D=2.0$.
(d) The jammed state at $F_D=0.0$ after the ramp-down.
}
\label{fig:4}
\end{figure}

In Fig.~\ref{fig:4}(a), we show the laned state at $F_D = 1.0$
for the system from Fig.~\ref{fig:3}(a,b) with
$\rho = 0.208$, where the lanes are reduced in width due to the relatively
low number of particles available.
Figure~\ref{fig:4}(b) shows the
same system in the jammed state at $F_D = 0.0$ after the ramp-down.
For the $\rho=0.93$ system from Fig.~\ref{fig:3}(c,d),
Fig.~\ref{fig:4}(c) shows the laned state at
$F_D = 2.0$, and Fig.~\ref{fig:4}(d) shows the
jammed state at $F_D = 0.0$ after the ramp-down, where
a polycrystalline solid appears.

\section{Perpendicularly Driven Particles} 

\begin{figure}
\includegraphics[width=\columnwidth]{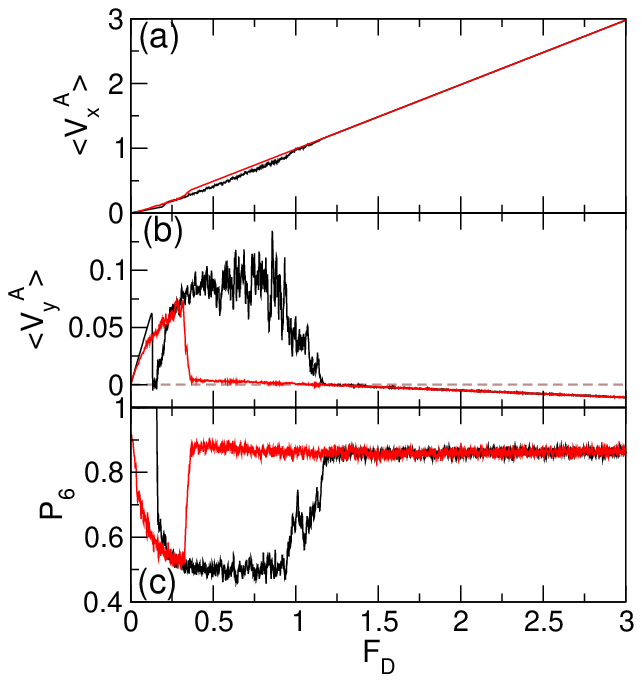}
\caption{A perpendicularly driven system with species A driven along $+x$
and species B driven along $+y$ at $\rho=0.441$.
Black curves are for ramp-up and red curves
are for ramp-down.
(a) $\langle V^A_x \rangle$ vs $F_D$.
(b) $\langle V^A_y \rangle$ vs $F_D$. The brown dashed line is drawn
at zero velocity to indicate the point at which the $y$ velocity goes negative.
(c) $P_6$ vs $F_D$.
There are four phases: a locked phase,
a one-dimensional (1D) decoupled phase,
a disordered flow phase,
and a high-drive tilted lane state.
}
\label{fig:5}
\end{figure}

\begin{figure}
\includegraphics[width=\columnwidth]{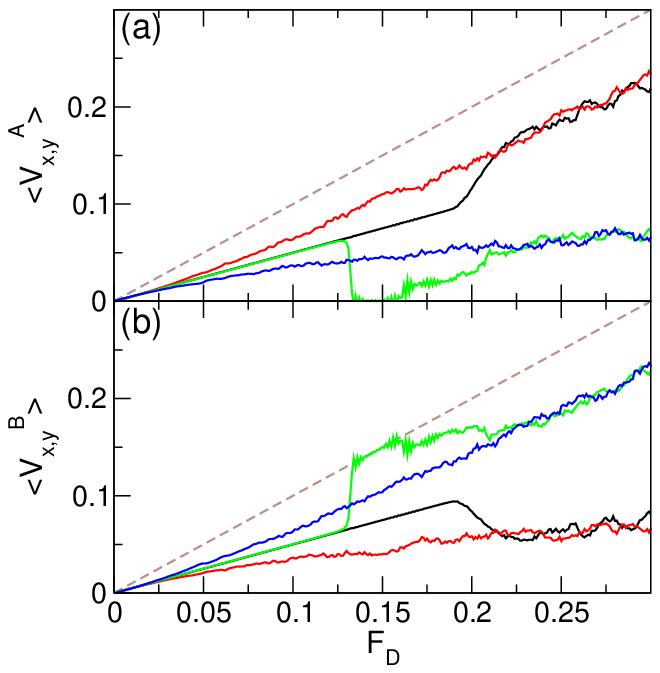}
\caption{The perpendicularly driven system at $\rho=0.441$ from
Fig.~\ref{fig:5}.
(a) Black and red: $\langle V_x^A\rangle$ vs $F_D$ during ramp-up and
ramp-down, respectively.
Blue and green: $\langle V_y^A\rangle$ vs $F_D$ during ramp-up and
ramp-down, respectively.
(b) Black and red: $\langle V_x^B\rangle$ vs $F_D$ during ramp-up and
ramp-down, respectively.
Blue and green: $\langle V_y^B\rangle$ vs $F_D$ during ramp-up and
ramp-down, respectively.
The data is shown only up to $F_D = 0.3$.
At low driving forces, there is a locked phase where
species A and B move together.
This is followed by a decoupled 1D flow phase,
where species A moves along $+x$ 
and species B moves along both $+x$ and $+y$.
Finally, there is a disordered flow regime at larger drives.
}
\label{fig:6}
\end{figure}

We next consider perpendicularly driven systems where
species $A$ is driven in the positive $x$-direction and species $B$
is driven in the positive $y$-direction.
In Fig.~\ref{fig:5}(a,b), we plot
$\langle V^A_x \rangle$ and $\langle V^A_y \rangle$,
respectively, versus $F_D$
for the ramp-up and ramp-down.
Figure~\ref{fig:5}(c) shows the
corresponding $P_6$ versus $F_D$ curves.
We identify four dynamic phases.
For $F_D < 0.129$, the system is in a jammed phase,
but unlike the oppositely driven system,
even though the particles are locked together,
they move as a rigid solid along $45^\circ$.
To show this more clearly,
in Fig.~\ref{fig:6}(a) we plot
$\langle V_x^A \rangle$ and $\langle V_y^A\rangle$ versus
$F_D$ during both the ramp-up
and ramp-down stages only up to $F_D=0.3$.
Figure~\ref{fig:6}(b) shows the same for the $B$ species, where we plot
$\langle V_x^B\rangle$ and $\langle V_y^B\rangle$ versus $F_D$
during ramp-up and ramp-down.
The dashed line in each panel is the velocity response
that would be expected for free particles interacting only with a
drive of $F_D$.
There is a coupling of $\langle V_x^A\rangle$ and $\langle V_x^B\rangle$
during the ramp-up and the two curves increase together with increasing
$F_D$ until $F_D=0.129$, at which point species A begins to move more
rapidly and species B begins to move more slowly
along $+x$.
Similarly, during the ramp-down, 
$\langle V_y^A\rangle$ and $\langle V_y^B\rangle$ undergo a recoupling
transition when $F_D$ drops below $F_D=0.125$; below this drive,
species A and B move at the same speed along $+y$.
Unlike the oppositely driven system,
at low drives in the perpendicularly driven system
the net particle velocities do not cancel, so that the jammed particle
assembly can undergo a net drift;
however, the drift speed is half the speed that would be expected
for noninteracting or free particles subjected to the same drive.

\begin{figure}
\includegraphics[width=\columnwidth]{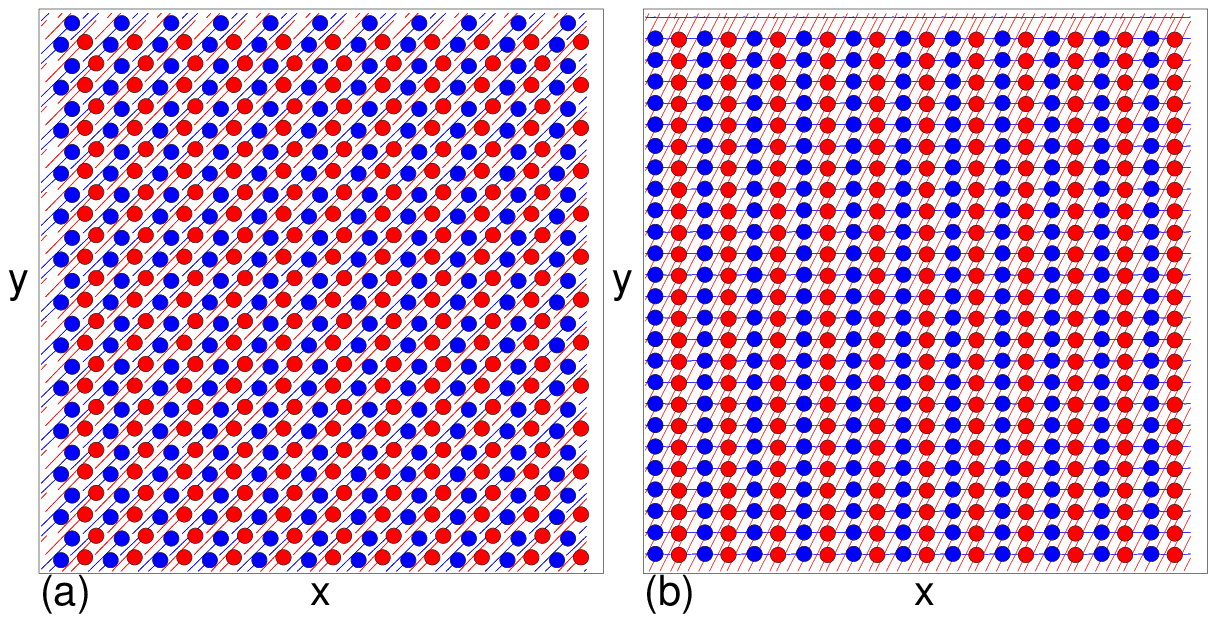}
\caption{Particle positions and trajectories for the perpendicularly
driven system with $\rho=0.441$
from Fig.~\ref{fig:5} where species A (blue) is driven along $+x$ and
species B (red) is driven along $+y$.
(a) The locked state at $F_D = 0.1$ where both particle species
move together as a rigid solid along 45$^\circ$.
(b) At $F_D = 0.15$,
species A is moving in the $+x$ direction
and species B is moving in both the $+x$ and $+y$ directions.
}
\label{fig:7}
\end{figure}

\begin{figure}
\includegraphics[width=\columnwidth]{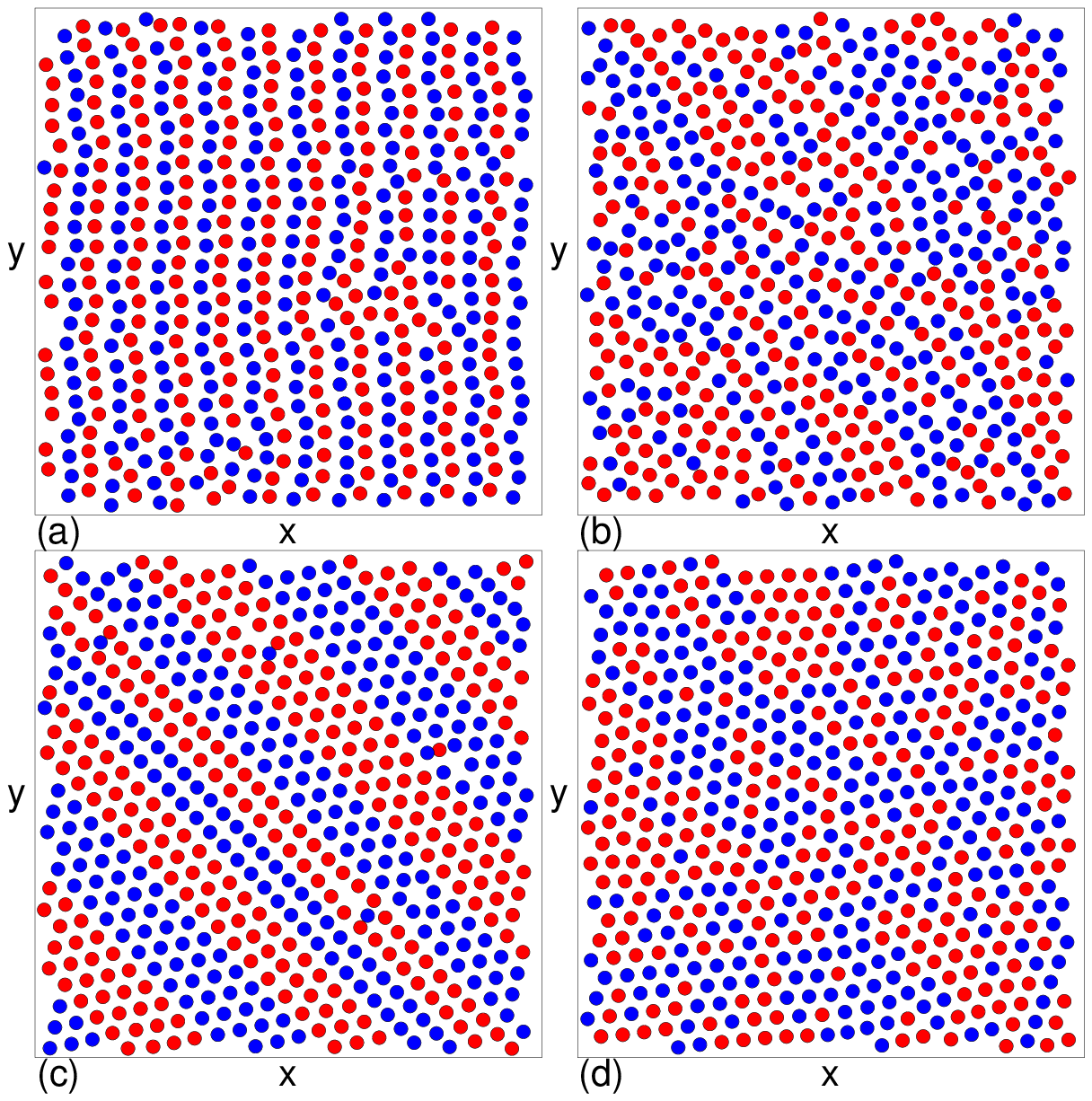}
\caption{Particle positions for the perpendicularly driven
system with $\rho=0.441$ from Fig.~\ref{fig:5} where species A (blue) is
driven along $+x$ and species B (red) is driven along $+y$.
(a) $F_D = 0.48$, just before the transition to the disordered phase.
(b) $F_D = 0.7$ in the disordered flow phase.
(c) The tilted lanes state at $F_D = 2.0$.
(d) The jammed state at $F_D=0.0$ after the ramp-down.
}
\label{fig:8}
\end{figure}

In Fig.~\ref{fig:7}(a), we show the particle positions and trajectories
at $F_D = 0.1$ for the
system from Fig.~\ref{fig:6},
where the particles are all moving as a rigid solid
along 45$^\circ$.
For $0.129 < F_D < 0.165$, there is a partial decoupling of the motion, with
species A moving only along the x-direction, causing
$\langle V_y^A \rangle$ to drop to zero.
At the same time, $\langle V_x^A \rangle$
continues to increase linearly with increasing $F_D$.
For species B, this decoupling transition is accompanied by
a jump up in $\langle V_y^B \rangle$,
while $\langle V_x^B\rangle$ continues to increase linearly
with increasing $F_D$.
We illustrate the particle configurations and trajectories in this
partially decoupled state at $F_D=0.15$ in
Fig.~\ref{fig:7}(b). The particles have formed well defined vertical
columns, and species A travels almost entirely along the
$+x$ direction while species B moves along both $+x$ and $+y$.
For $0.165 \leq F_D < 0.21$, species A picks up some finite motion
along the $+y$-direction,
but the overall motion remains nearly 1D.
We label these states ``locked'' where the two species move together and
``1D flow'' where there is a partial decoupling of the motion.
At $F_D=0.21$, the motion becomes disordered,
which correlates with a jump up in
$\langle V^A_x\rangle$ and $\langle V^A_y \rangle$
and a drop in $\langle V^B_x\rangle$ and $\langle V^B_y \rangle$.
The velocity signals exhibit stronger fluctuations
above the disordering transition.
In Fig.~\ref{fig:8}(a), we show the particle configurations at $F_D=0.48$,
just before the onset of the disordered phase,
where the motion is partially one-dimensional.
Figure~\ref{fig:8}(b) shows the particle configurations
in the disordered phase at $F_D = 0.7$.
In Fig.~\ref{fig:5}(c), $P_6 = 1.0$
in both the jammed phase and the 1D motion phase, but $P_6$ drops
in the disordered flow phase.
Even though species A is not driven in the $y$-direction,
Fig.~\ref{fig:5}(b) indicates that
there is still a finite velocity component
$\langle V_y^A\rangle$ in the $y$-direction, produced due to interactions
with the species B particles that are
driven in the $y$-direction.
For $F_D > 1.2$, the system forms an ordered tilted laned state,
which is accompanied by a drop in both
$\langle V^A_x \rangle$ and $\langle V^A_y \rangle$ in
Fig.~\ref{fig:5}(a,b).
At the onset of the tilted laned state,illustrated in Fig.~\ref{fig:8}(c)
at $F_D=2.0$, 
$P_6$ increases to $P_6\approx 0.89$
since the system becomes more ordered.
Within the laned state,
$\langle V^A_x \rangle$ and $\langle V^B_y \rangle$ reach values
close to the free particle limit, and both quantities
increase linearly with increasing $F_D$.

\begin{figure}
\includegraphics[width=\columnwidth]{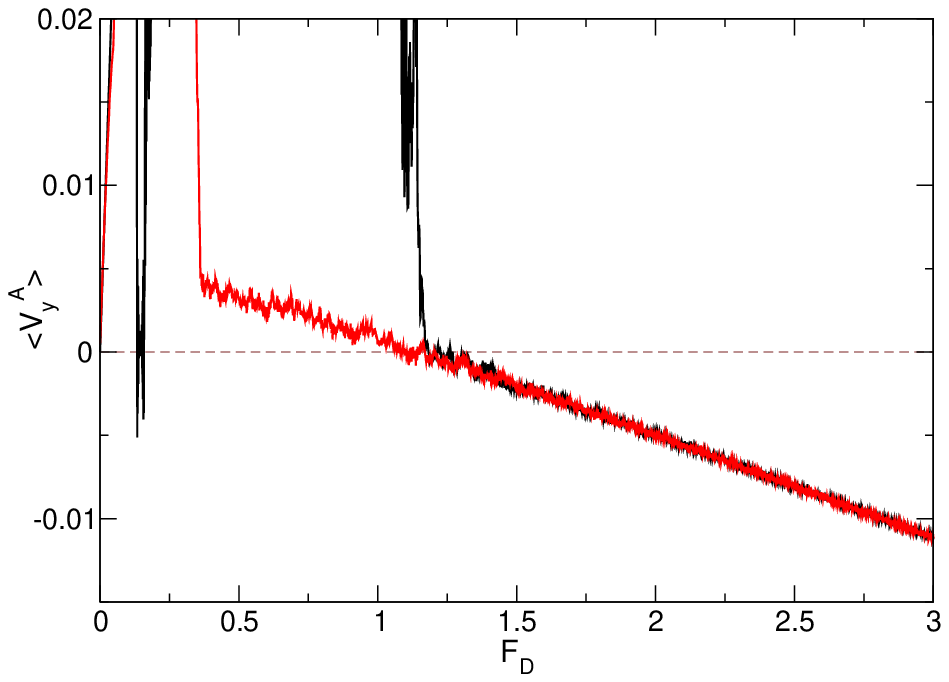}
\caption{$\langle V_y^A\rangle$ vs $F_D$ for the perpendicularly driven system
from Fig.~\ref{fig:5} during ramp-up (black) and ramp-down (red).
A negative drag effect occurs at higher drives where
$\langle V_y^A \rangle < 0.0$. In the negative drag regime,
species A is moving in the direction
opposite to the driving direction of
species B.
}
\label{fig:9}
\end{figure}

We find that
within the tilted lane state,
$\langle V^A_y \rangle$ has an extended region of negative velocity that
is visible
in Fig.~\ref{fig:5}(b).
This is illustrated more clearly in the zoomed in plot
of $\langle V_y^A\rangle$ versus $F_D$ for a portion of the ramp-up and
ramp-down shown in Fig.~\ref{fig:9}.
Here species B, which is being driven along $+y$,
creates a negative drag effect on species A,
causing species A to move in the opposite or $-y$ direction
instead of being dragged along by species B.
The negative $\langle V^A_y \rangle$ regime arises as a result
of the tilting of the laned state
seen in Fig.~\ref{fig:8}(c) at $F_D = 2.0$.
The species A particles
are driven along $+x$, but when they encounter the species B particles,
which have assembled into a tilted wall,
the flow of species A is partially blocked by species B.
Upon reaching the tilted barrier, the species A particles
are slightly deflected in the $-y$ direction,
leading to a negative $y$-velocity and a
positive $x$-velocity for species A.
When the drive is in the ramp-down regime,
the tilted laned state persists
down to a drive of $F_D=0.35$,
which is about 3.5 times smaller than the drive at which the laned state
appeared during the ramp-up, indicating the presence of strong
hysteresis.
The system re-enters a disordered state during ramp-down
but does not form the same 1D motion and
ordered locked states that appeared during ramp-up.
Instead, the ramp-down produces
mostly triangular solids composed of mixed species, as illustrated
in Fig.~\ref{fig:8}(d) at $F_D = 0.0$ after the ramp-down.
The locked state and the 1D state that appeared during ramp-up preserve
a memory of how the system was initially prepared.
If we ramp-up the system for a second time after the original ramp-up and
ramp-down cycle, the locked state and 1D flow state are both lost and
replaced by a disordered state and a laned state. Even during the
second ramp-up, there is still strong hysteresis
across the transition to the tilted lanes state.
The formation of tilted lanes has been observed
in previous work on oppositely driven particles when a rule is introduced
that causes particles to move preferentially to one side when interacting
with an oppositely moving particle \cite{Bacik23}. 
In our case, the tilt also occurs due to the particles passing each
other on a preferred side, but the preference is produced by
the fact that the species B particles are driven in the $+y$ direction.
The angle of the tilt depends on the ratio of $F^A_D/F^B_D$,
which we vary in the next section.
If we apply a negative $y$ direction drive
to species B, the tilt of the lanes is reversed.

\begin{figure}
\includegraphics[width=\columnwidth]{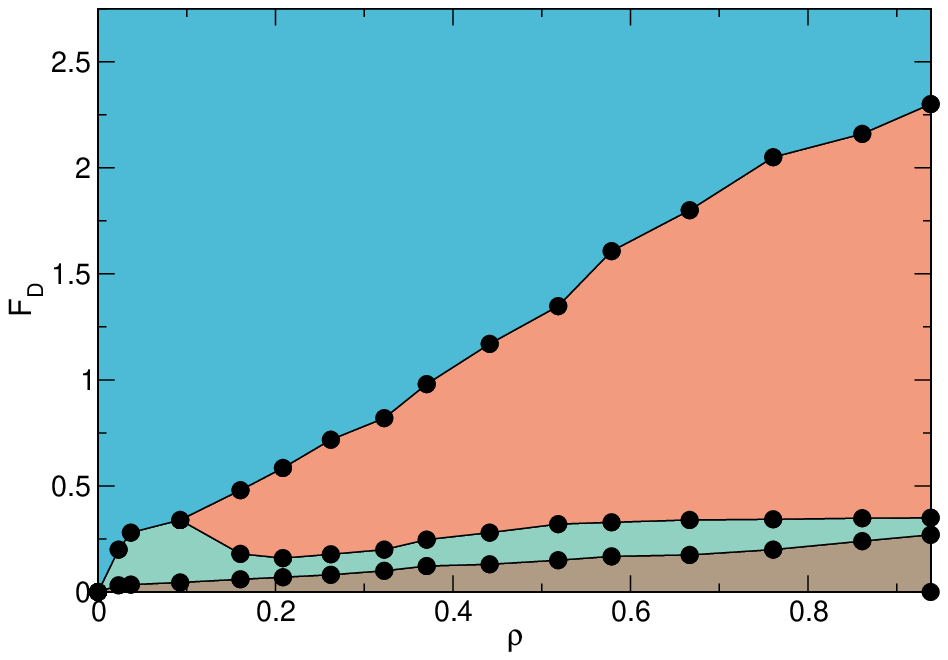}
\caption{Dynamic phase diagram as a function of $F_D$ vs $\rho$
for the perpendicularly driven system from Fig.~\ref{fig:5}.
The following phases are present:
locked (brown),
1D flow (green),
disordered flow (orange),
and tilted lanes (blue).
}
\label{fig:10}
\end{figure}

In Fig.~\ref{fig:10}, we plot a dynamic phase diagram as a function
of $F_D$ versus $\rho$ for the perpendicularly driven system from
Fig.~\ref{fig:5}, where we highlight the locked state,
the 1D flow state, the disordered flow phase,
and the tilted laned phase.
As $\rho$ increases, 
the locked phase grows in extent due to the increased interaction between
the more closely spaced particles.
The disordered state also increases in extent with increasing $\rho$.
When $\rho < 0.1$, there is no disordered flow regime.

\begin{figure}
\includegraphics[width=\columnwidth]{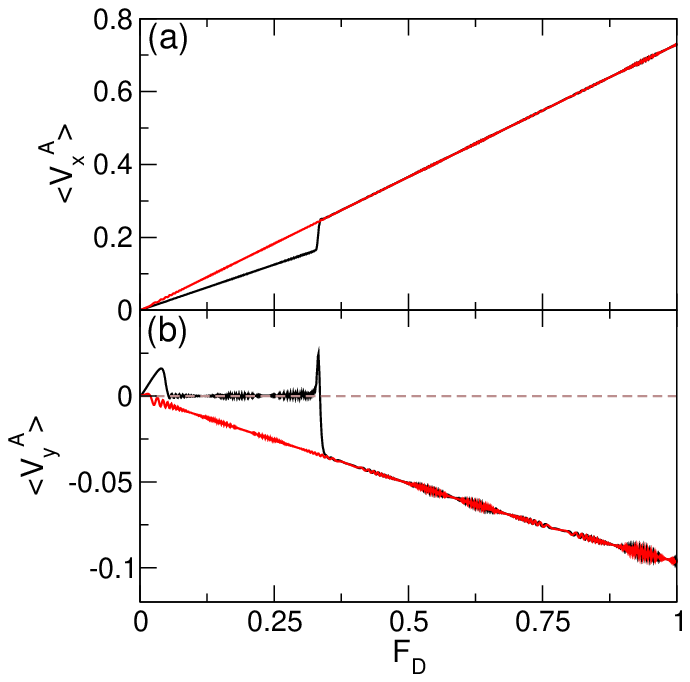}
\caption{(a) $\langle V^{A}_{x}\rangle$ and
(b) $\langle V^{A}_{y}\rangle$ vs $F_{D}$ for
the perpendicularly driven system from Fig.~\ref{fig:10} with $\rho = 0.093$
during ramp-up (black) and ramp-down (red). 
}
\label{fig:11}
\end{figure}

\begin{figure}
\includegraphics[width=\columnwidth]{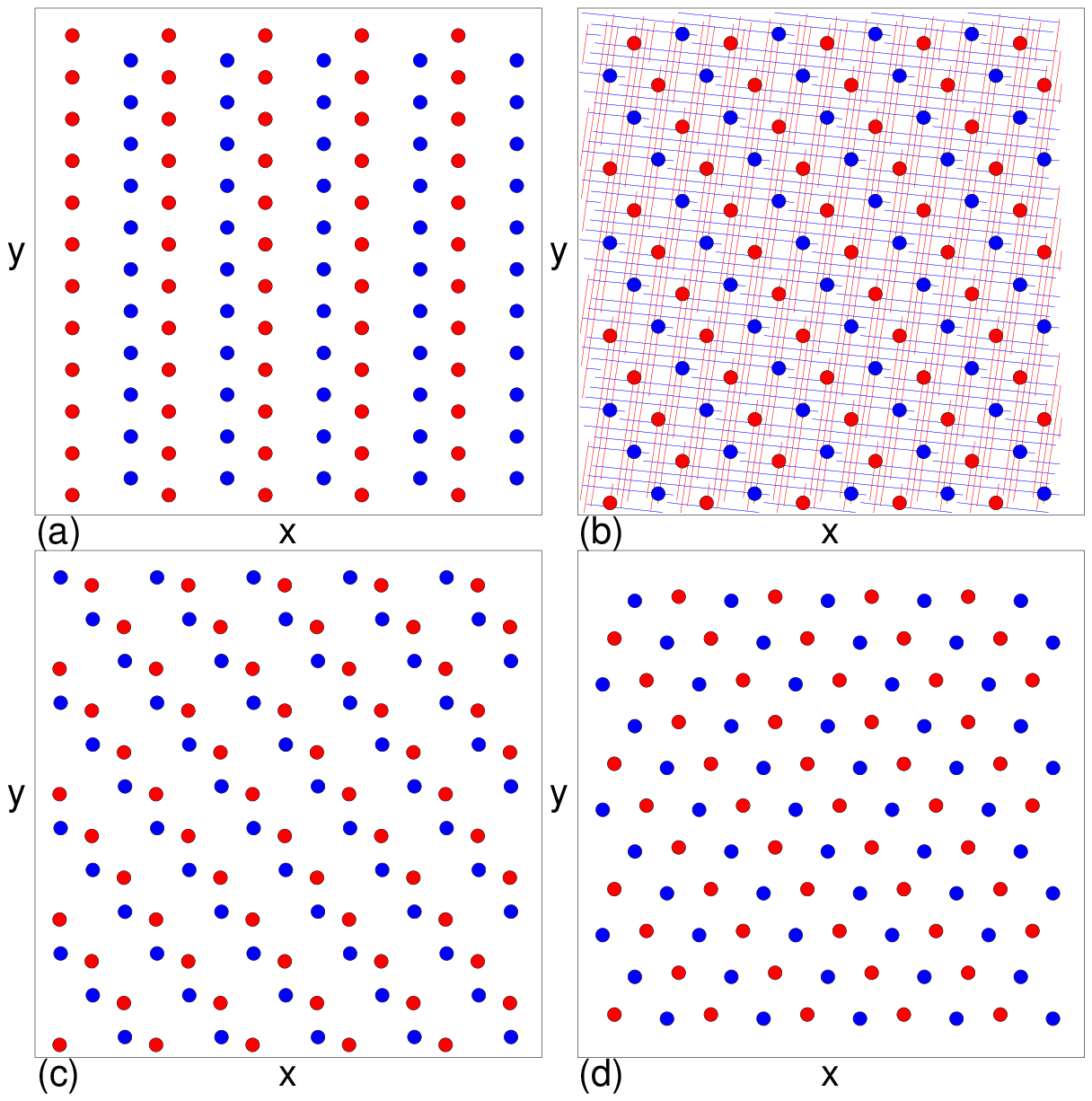}  
\caption{Particle positions for the perpendicularly driven system with
$\rho=0.093$ from Fig.~\ref{fig:11} where species A (blue) is driven along
$+x$ and species B (red) is driven along $+y$.
(a) The 1D state at $F_D = 0.3$.
(b) Particle positions and trajectories in the tilted lanes
state at $F_D = 0.6$.
(c) The compressed tilted lanes state at $F_D = 2.5$.
(d) The pinned triangular solid at $F_D = 0.0$ after the ramp-down.
} 
\label{fig:12}
\end{figure}

In Fig.~\ref{fig:11}(a,b), we plot
$\langle V^A_x\rangle$ and $\langle V^A_y\rangle$ versus
$F_D$ for the system from
Fig.~\ref{fig:10} at $\rho = 0.093$ during
the ramp-up and ramp-down,
showing the absence of a disordered flow phase
and the presence of extended regions of 1D-like motion.
We show the particle positions for the
$\rho=0.093$ system 
in Fig.~\ref{fig:12}(a) at$F_D = 0.3$,
where species A is moving only in the $+x$ direction and
species B is moving in the $+y$ and $+x$ directions.
The particles assemble into stripes that are
aligned in the $y$-direction.
At $F_D=0.6$ in Fig.~\ref{fig:12}(b), where we plot both positions and
trajectories of the particles,
the system is in the laned state.
Here species A is moving mostly along $+x$ and slightly along $-y$,
while species B is moving along $+y$ and $+x$.
Fig.~\ref{fig:12}(c) shows that at $F_D = 2.5$, the lanes are
compressed, forming a tilted stripe-like pattern
as species A pushes against species B.
In Fig.~\ref{fig:12}(d), an
ordered laned state appears
at $F_D = 0.0$ after the ramp-down.

\section{Fixed Longitudinal Drive and Increasing Perpendicular Drive}

\begin{figure}
\includegraphics[width=\columnwidth]{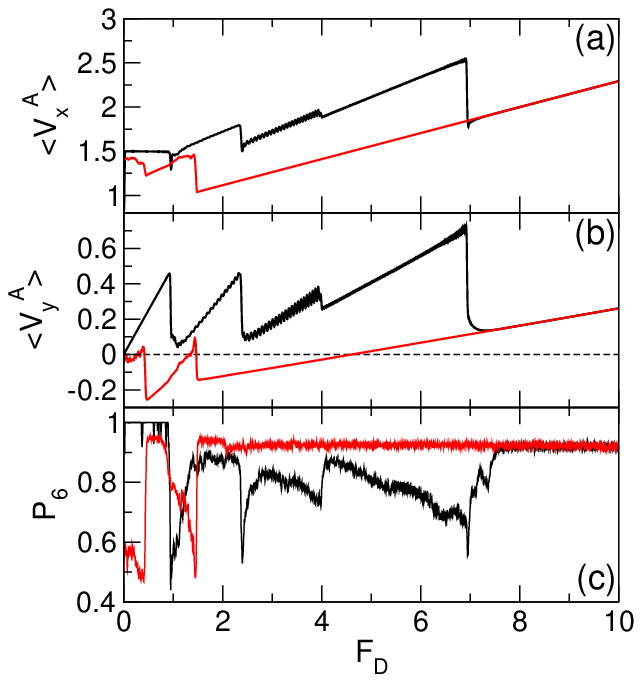}
\caption{A perpendicularly driven system with $\rho=0.44$ where
species A is driven in the $+x$ direction
at $F_D^A=1.5$ and species B is driven along $+y$ at $F_D$.
Black curves are for the ramp-up and red curves are for the ramp-down
of $F_D$.
(a) $\langle V^A_x \rangle$ vs $F_D$.
(b) $\langle V^A_y \rangle$ vs $F_D$.
(c) $P_6$ vs $F_D$.
}
\label{fig:13}
\end{figure}

\begin{figure}
\includegraphics[width=\columnwidth]{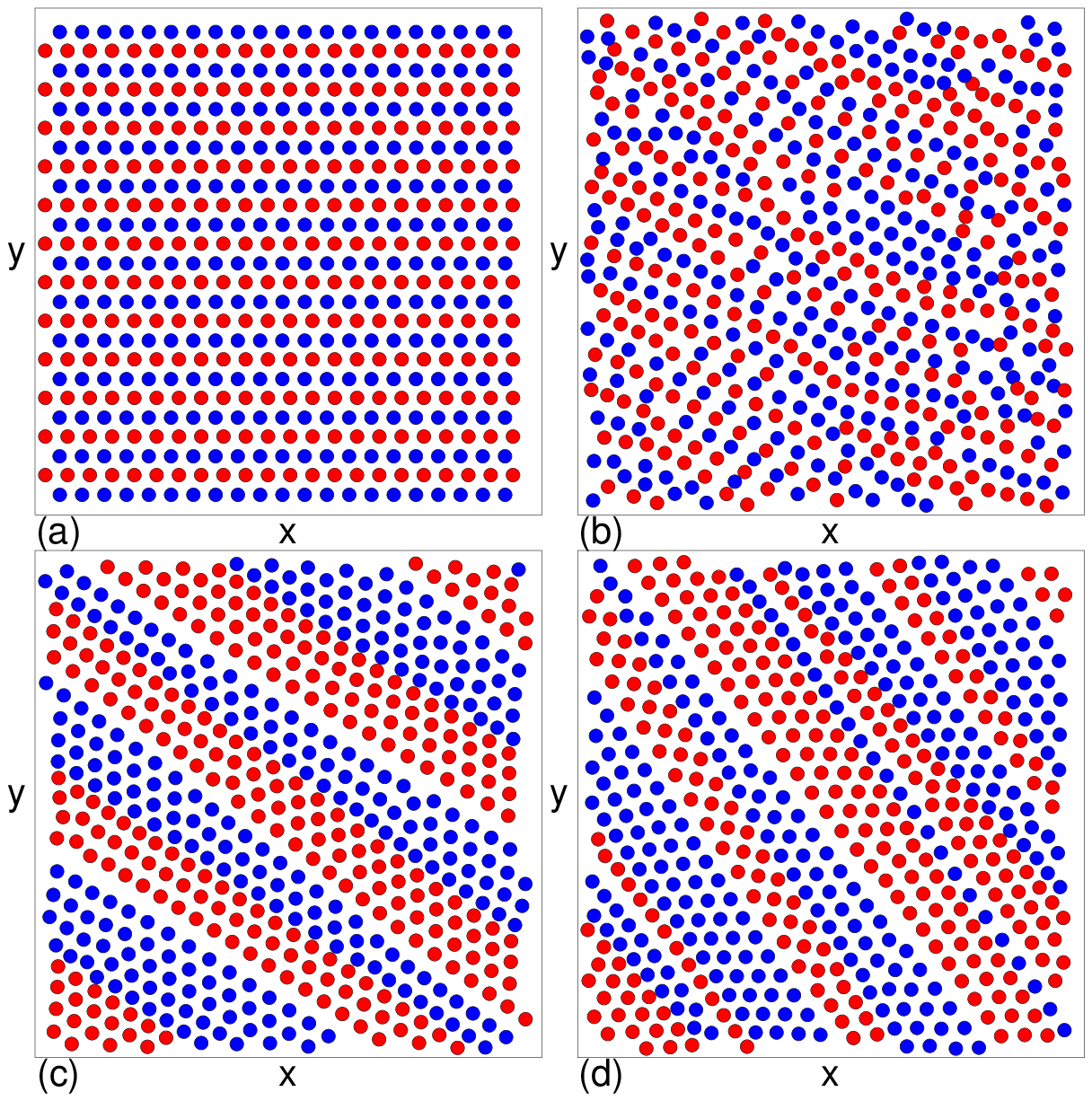}
\caption{Particle configurations for the perpendicularly driven
system with $\rho=0.44$ from Fig.~\ref{fig:13} where species A (blue) is driven at
$F_D^A=1.5$ along $+x$ and species B (red) is driven along $+y$ at
varied $F_D$.
(a) The 1D flow phase at $F_D = 0.5$.
(b) Disordered flow at $F_D = 1.05$.
(c) Tilted lanes state at $F_D = 2.0$.
(d) Tilted lanes state at $F_D = 3.0$.
}
\label{fig:14}
\end{figure}

We next hold the species A drive fixed in the $x$ direction at a value
$F_D^A=1.5$ while we sweep the $y$ direction species B
drive $F_D^B=F_D$ up and down from $F_D=0.0$.
In Fig.~\ref{fig:13}(a,b) we plot
$\langle V^A_x \rangle$
and $\langle V^A_y \rangle$,
respectively, versus
$F_D$ for a system with $\rho=0.44$, while
Fig.~\ref{fig:13}(c) shows the corresponding $P_6$ versus
$F_D$ curve.
At $F_D=0$ in Fig.~\ref{fig:13}(a),
$\langle V^A_x \rangle$ initially has a value close to
$\langle V^A_x\rangle = F_D^A=1.5$, indicating
that the motion of species A is decoupled from
that of species B.
Additionally, $P_6$ is close to one
since the system forms an ordered crystal.
As $F_D$ continues to increase,
there is a very small
locked phase just above $F_D=0.0$ where
species A and B move together.
There is then a transition to 
1D flow, where
$\langle V^A_y \rangle$ increases linearly with increasing $F_D$.
The 1D flow persists
up to $F_D=0.9$.
In Fig.~\ref{fig:14}(a), we show the
particle configurations in the 1D flow phase at
$F_D = 0.5$,
where species A forms 1D chains aligned in the $x$-direction
that can easily move between the species B particles.
At $F_D = 0.9$, there is a sharp drop in
$\langle V^A_y \rangle$ accompanied by a drop in $P_6$ to
$P_6=0.5$ 
and the system becomes disordered, as shown in
Fig.~\ref{fig:14}(b) at $F_D = 1.05$.
After this disordering transition,
$\langle V_x^A \rangle$ increases linearly
with increasing $F_D$, and $P_6$
also increases up to $P_6=0.85$ when
a tilted lane state appears,
as shown in Fig.~\ref{fig:14}(c) at $F_D = 2.0$,
where the tilt angle is approximately 30$^\circ$ from the $x$ axis.

\begin{figure}
\includegraphics[width=\columnwidth]{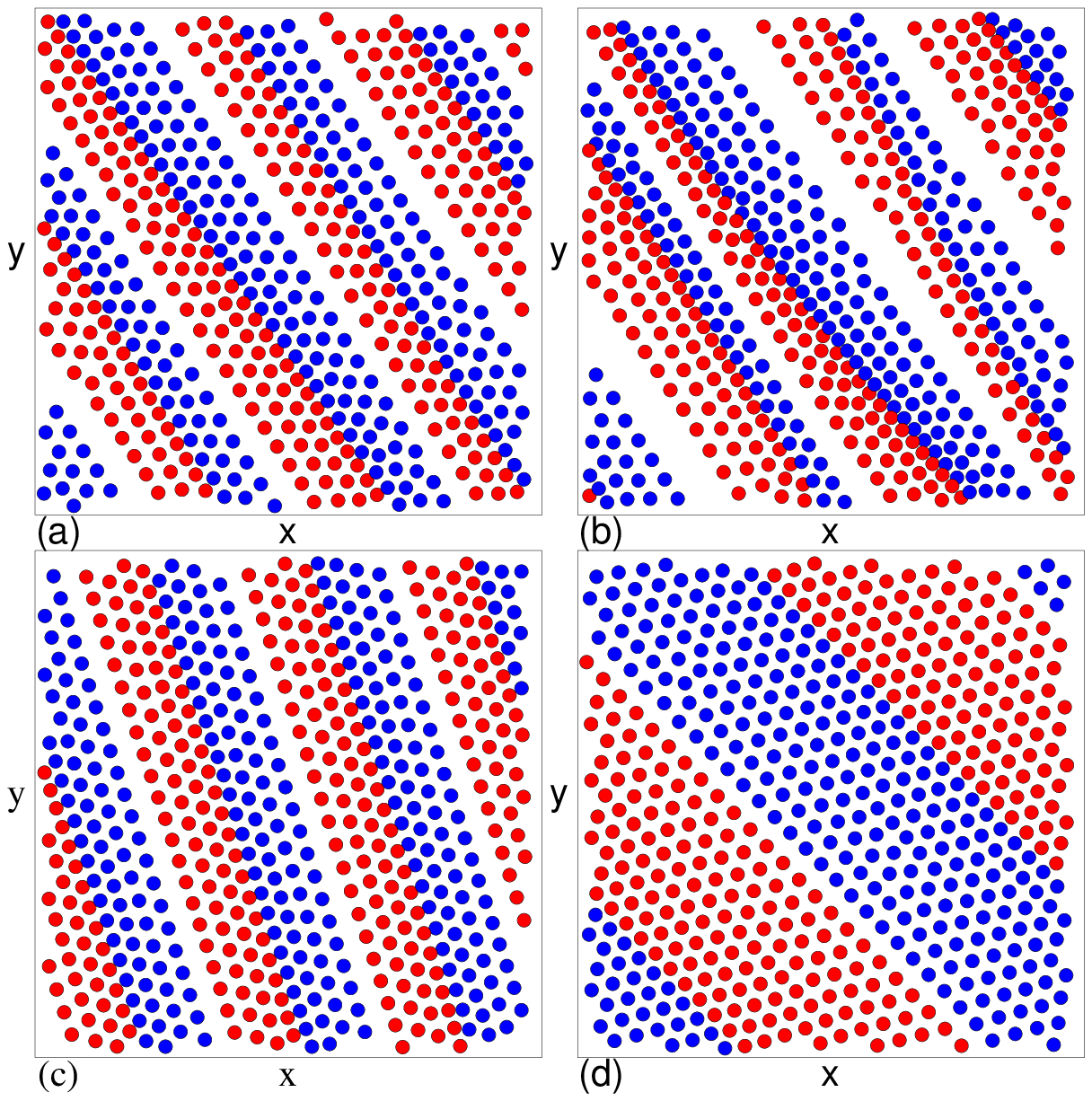}
\caption{Particle configurations for the perpendicularly driven system
from Fig.~\ref{fig:13} with $\rho=0.44$ where species A (blue) is driven at $F_D^A=1.5$
along $+x$ and species B (red) is driven along $+y$ at varied $F_D$.
(a) Tilted lanes state at $F_D = 5.4$.
(b) Striped tilted state at $F_D = 6.82$.
(c) Tilted lanes state at $F_D = 8.1$.
(d) Tilted lanes state at $F_D = 0.6$ during the ramp-down.
}
\label{fig:15}
\end{figure}

Near $F_D = 2.4$ in Fig.~\ref{fig:13}, there is another
disordering event marked
by simultaneous drops in $P_6$, $\langle V_x^A \rangle$,
and $\langle V^A_y \rangle$.
After this second disordering event,
a new tilted lane state forms,
as illustrated in Fig.~\ref{fig:14}(d) at $F_D = 3.0$.
The tilt of the lanes has increased to a value
close to 60$^\circ$, indicating that as $F_D$ increases,
the system is attempting to match the motion of the laning particles
to the net direction of the external drive.
As $F_D$ increases further,
a third disordering transition occurs and is followed by the
appearance of yet another laned state, shown in
Fig.~\ref{fig:15}(a) at $F_D = 5.4$.
The particles have formed
a more stripe-like structure
with increased compression.
In this regime,
both $\langle V_x^A \rangle$ and $\langle V_y^A \rangle$ increase linearly
with increasing $F_D$,
and the laned state from Fig.~\ref{fig:15}(a)
becomes increasingly compressed until it reaches a
maximally compressed state,
shown at $F_D = 6.82$ in Fig.~\ref{fig:15}(b).
At $F_D=7.0$, there is a fourth
disordering transition along with
large drops
in $\langle V_x^A \rangle$ and $\langle V_y^A\rangle$.
A new laned state with $P_6=0.92$ emerges at higher drives,
as shown in Fig.~\ref{fig:15}(c) at $F_D = 8.1$.
The lanes have become much wider, the compression is smaller,
and the tilt of the lanes is even greater.
We begin to decrease $F_D$ again after reaching
$F_D = 10.0$, and
find that the system remains in the same laned state shown
in Fig.~\ref{fig:15}(c) all the way back down to $F_D = 1.5$,
where a disordering transition occurs.
As the drive continues to decrease,
there is a jump up in $\langle V^A_x \rangle$
and $\langle V^A_y \rangle$
when a new laned state with a reduced tilt emerges,
as shown in Fig.~\ref{fig:15}(d) at $F_D = 0.6$.
The system becomes disordered again at low $F_D$ when
the particles attempt to form a laned state aligned
with the $x$ direction.
This result indicates that there is very strong
hysteresis across the transitions among the different tilted lane states.

\begin{figure}
\includegraphics[width=\columnwidth]{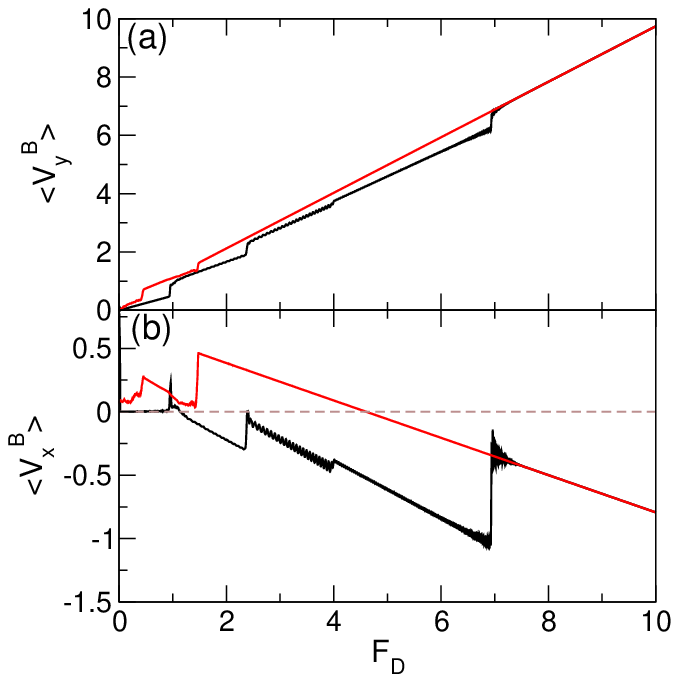}
\caption{The perpendicularly driven system from Fig.~\ref{fig:13}
with $\rho=0.44$ where
species A is driven at $F_D^A=1.5$ along $+x$ and species B is driven
along $+y$ at $F_D$.  
Black curves are for the ramp-up and red curves are for the ramp-down
of $F_D$.
(a)  $\langle V^B_y \rangle$ vs $F_D$.
(b) $\langle V^B_x \rangle$ vs $F_D$.
}
\label{fig:16}
\end{figure}

In Fig.~\ref{fig:16}(a,b), we
plot $\langle V^B_y \rangle$ and $\langle V^B_x \rangle$,
respectively, for the system in Fig.~\ref{fig:13}.
The transitions among the different tilted states appear as
jumps up in $\langle V^B_y \rangle$,
and species B has a higher $y$ velocity during the ramp-down than
on the ramp-up.
The velocity increase across the transitions
is consistent with
the fact that the tilt of the lanes
becomes more aligned with
the $y$-direction drive after each transition.
In $\langle V^B_x \rangle$ we find an extended region of negative velocity,
where species B is moving in the direction opposite to the driving
force applied to species A. This is the result of a guidance effect produced
by the tilted lanes.

As the driving force $F_D$ in the $y$-direction increases,
the system attempts
to form a laned state that reduces the
frequency of particle-particle collisions.
At low $F_D$, this is best accomplished by forming particle lanes that
are aligned along the $x$ direction,
but as $F_D$ increases, the optimal lane
orientation rotates to match
the net direction of the applied driving forces, which becomes closer and closer
to the $y$ direction.
Once a laned state has formed that matches a particular
value of $F_D$, the lanes
become increasingly compressed as $F_D$ further increases.
When the energy cost of this compression becomes too high, the lanes
break apart
into a disordered state that
can then reform into a new tilted lane state that is
better aligned with the current net driving force direction.
As $F_D$ is swept up from zero,
this process of lane formation and destruction occurs repeatedly,
leading to a series of disordering transitions and a succession of
tilted states with tilt angles that become closer to 90$^\circ$ as
$F_D$ increases.
In Fig.~\ref{fig:13}, we sweep the drive up to a maximum value
of $F_D = 10$.
For higher drives, there could be additional disordering transitions and
additional tilted lane states.
The drops in $\langle V^A_x \rangle$ and increases
in $\langle V^B_y \rangle$ occur whenever
a new laned state forms that has a tilt which is more closely
aligned with the $y$-direction. This
leads to an increase in the $y$-direction flow of species B
and increased $x$-direction drag on species A.
Once we begin to decrease $F_D$ instead of increasing it,
the laned state
shown in Fig.~\ref{fig:15}(c) is subjected to decompression rather than
compression. The laned configuration is stable against decompression, so
the system remains in the same laned state down to $F_D=F_D^A=1.5$.
As $F_D$ drops below $F_D^A$, the balance of forces is lost and the
lanes begin to compress with decreasing $F_D$. This results in a
disordering transition followed by formation of a new laned
state that is tilted closer to
the $x$ direction, as shown in Fig.~\ref{fig:15}(d) at
$F_D=0.6$ during the ramp-down.

\begin{figure}
\includegraphics[width=\columnwidth]{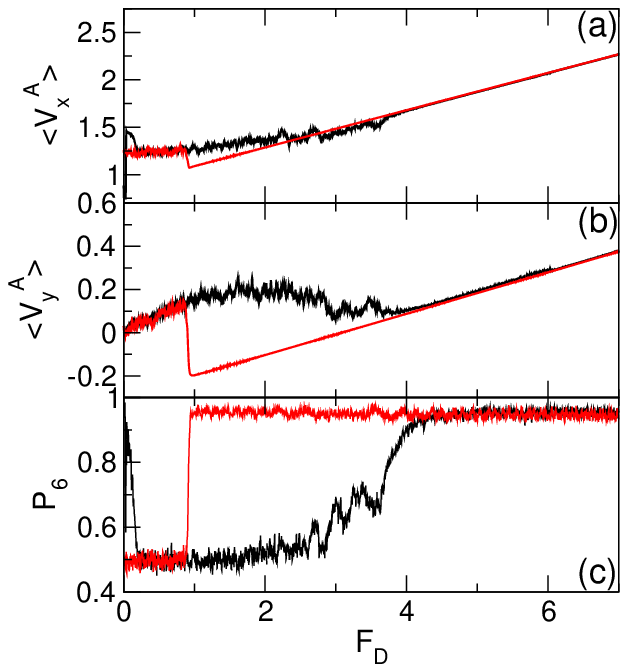}
\caption{A perpendicularly driven system with $\rho=0.94$ where
species A is driven at $F_D^A=1.5$ along $+x$ and species B is driven
along $+y$ at $F_D$.
Black (red) curves are for the ramp-up (ramp-down) of $F_D$.
(a) $\langle V^A_x \rangle$ vs $F_D$.
(b) $\langle V^A_y \rangle$ vs $F_D$.
(c) $P_6$ vs $F_D$.
}
\label{fig:17}
\end{figure}

\begin{figure}
\includegraphics[width=\columnwidth]{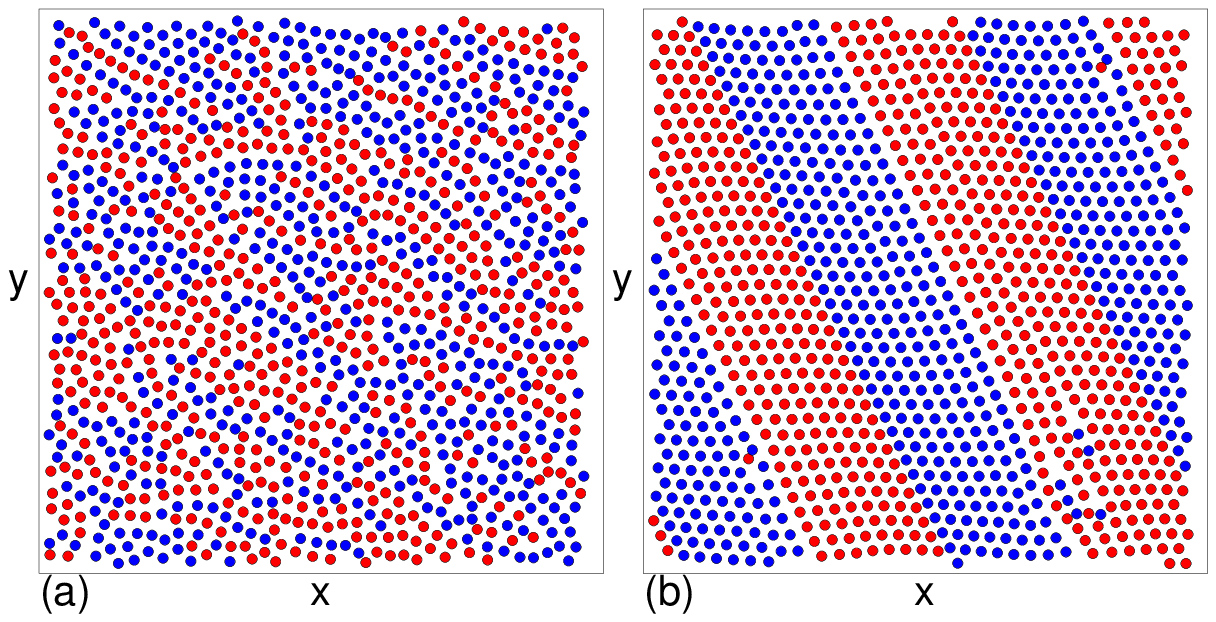}
\caption{Particle configurations for the perpendicularly driven
system from Fig.~\ref{fig:17} with $\rho=0.94$ where
species A (blue) is driven at $F_D^A=1.5$ along $+x$ and species B (red)
is driven along $+y$ at varied $F_D$.
(a) The disordered tilted flow state at $F_D = 2.0$.
(b) The tilted laned state at $F_D = 6.0$.
}
\label{fig:18}
\end{figure}

When we vary the particle density for the same driving protocol,
we find that in general, increasing $\rho$ produces expanded regions
of disordered flow.
For example, in Fig.~\ref{fig:17}(a,b) we plot
$\langle V_x^A \rangle$ and $\langle V_y^A \rangle$,
respectively, versus $F_D$
at fixed $F_D^A = 1.5$ for a system with $\rho = 0.94$.
The initial ordered state persists up to
$F_D = 0.15$, at which point the particle motion becomes disordered.
A tilting of the disordered domains is visible in
the image of the particle positions shown in
Fig.~\ref{fig:18}(a) at $F_D = 2.0$.
As $F_D$ increases further,
the system orders into a laned state, shown
in Fig.~\ref{fig:18}(b) at $F_D = 6.0$.
In Fig.~\ref{fig:17}(c), we plot
the corresponding $P_6$ versus $F_D$, where we find
that the laned state persists during the ramp-down
until $F_D = 0.9$, at which point the system disorders.
At higher densities, we find
that there are fewer tilting transitions, and that the transitions that
do occur
are shifted to higher values of $F_D$.

\begin{figure}
\includegraphics[width=\columnwidth]{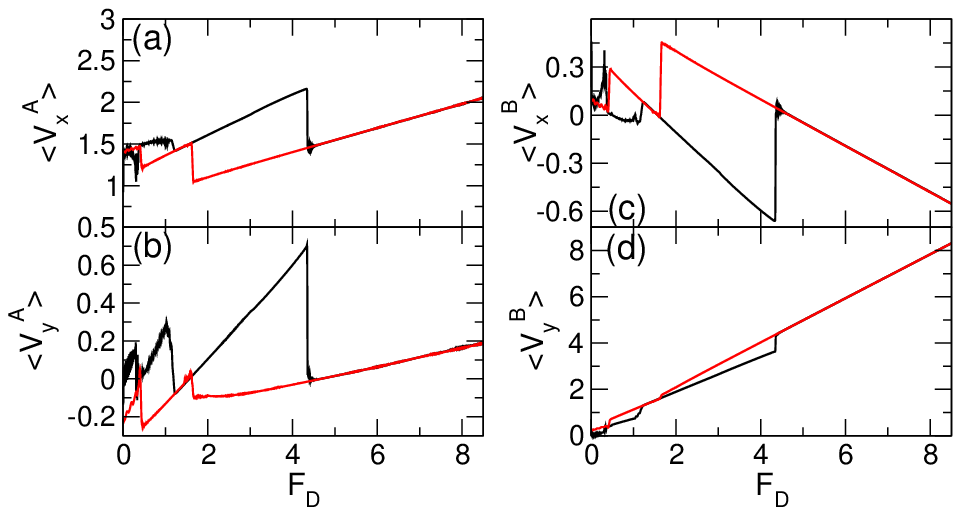}
\caption{A perpendicularly driven system with a low density of
$\rho=0.0925$ where species A is driven at $F_D^A=1.5$ along $+x$ and
species B is driven along $+y$ at $F_D$.
Black (red) curves are for the ramp-up (ramp-down) of $F_D$.
(a) $\langle V_x^A \rangle$ vs $F_D$.
(b) $\langle V_y^A \rangle$ vs $F_{D}$.
(c) $\langle V_x^B \rangle$ vs $F_D$.
(d) $\langle V_y^B \rangle$ vs $F_D$.
}
\label{fig:19}
\end{figure}

\begin{figure}
\includegraphics[width=\columnwidth]{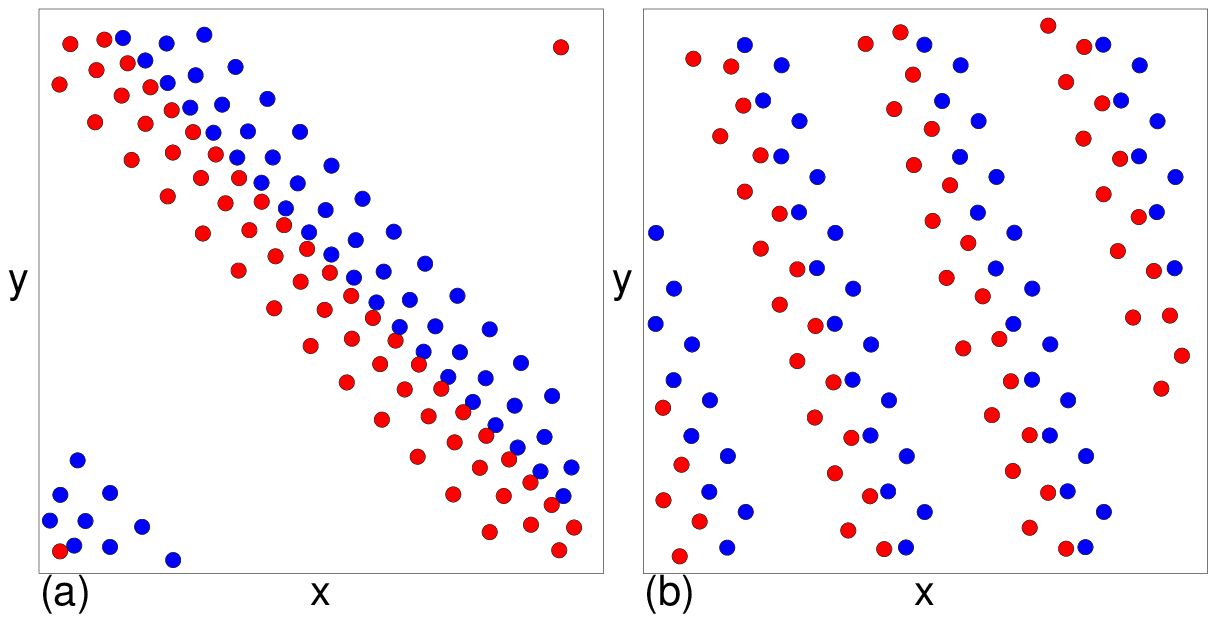}
\caption{(a) Particle configurations for the perpendicularly driven system
from Fig.~\ref{fig:19} with $\rho=0.0925$ where species A (blue) is driven
at $F^A_D=1.5$ along $+x$ and species B (red) is driven along $+y$ at
varied $F_D$.
(a) A single tilted stripe at $F_D = 3.0$.
(b) The next higher laned state at $F_D = 6.0$.}     
\label{fig:20}
\end{figure} 

In Fig.~\ref{fig:19}(a,b), we plot
$\langle V_x^A \rangle$ and
$\langle V_y^A \rangle$, respectively,
versus $F_{D}$ for a system with $F^{A}_{D} = 1.5$
at a low density of $\rho = 0.0925$.
Figure~\ref{fig:19}(c,d) shows the corresponding
$\langle V_x^B \rangle$ and $\langle V_y^B \rangle$,
respectively, versus $F_{D}$.
At this density, there is no disordered phase,
but there are transitions among
different ordered states at $F_{D} = 0.35$, $1.5$, and $4.4$.
These transitions are accompanied
by large changes in the velocities.
During the ramp-down,
there are two different hysteretic regimes,
but for $1.2 < F_{D} < 1.6$, the hysteresis is absent.
In Fig.~\ref{fig:20}(a), we show that
at $F_D = 3.0$, a single tilted stripe state appears.
Figure~\ref{fig:20}(b) shows the system at $F_D=6.0$
after the next laning transition has
occurred, where there are now two stripes tilted at a higher angle.

\begin{figure}
\includegraphics[width=\columnwidth]{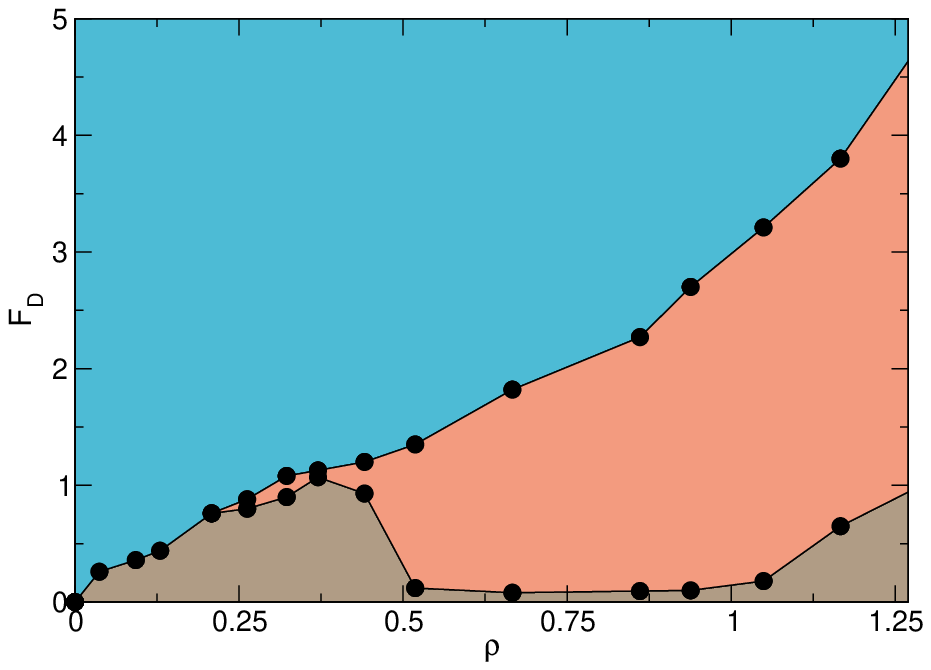}
\caption{Dynamic phase diagram as a function of
$F_D$ vs $\rho$ for a perpendicularly driven
system where species A is driven in the $+x$ direction
at fixed $F^A_D = 1.5$ and species B is driven in the $+y$
direction at varied $F_D$.
We find the following phases:
locked (brown),
disordered flow (orange),
and laned states (blue).
Here we do not distinguish between different laned states in the
laning regime, and we do not mark
the small disordered regions
that occur between the transitions among different laned states.}
\label{fig:21}
\end{figure}

By conducting a series of simulations, we can map out
the different phases, as shown in Fig.~\ref{fig:21}
as a function of $F_D$ versus $\rho$ for fixed $F^A_D = 1.5$.
We observe a locked phase where the motion of
species A is along the $x$-direction,
a disordered flow phase,
and tilted lane states, which include all
the different types of laned states.
For $\rho > 0.44$, the disordered flow regime increases in width,
and the drive at which a transition into the laned states occurs
increases with increasing $\rho$.
When $\rho < 0.2$, there is only a very small region of disordered
flow phase.
Within the laned states regime, there are
small regions of disordered flow
between the different laned states,
but these are not shown in the phase diagram.

\section{Varied Longitudinal Drive And Increasing Perpendicular Drive} 

\begin{figure}
\includegraphics[width=\columnwidth]{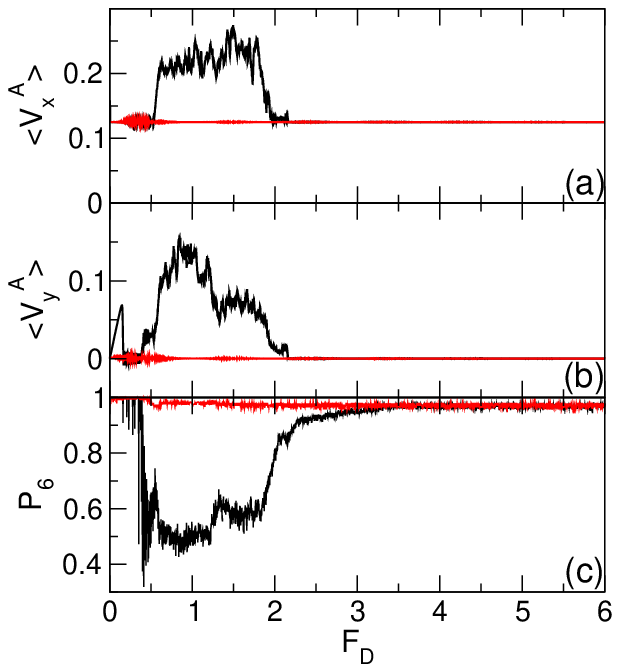}
\caption{A perpendicularly driven system with $\rho=0.44$ where species A
is driven at $F_D^A=0.25$ along $+x$ and species B is driven along $+y$ at
$F_D$.
Black (red) curves are for the ramp-up (ramp-down) of $F_D$.
(a) $\langle V_x^A \rangle$ vs $F_D$.
(b) $\langle V_y^A \rangle$  vs $F_D$.
(c) $P_6$ vs $F_D$.
}
\label{fig:22}
\end{figure}

\begin{figure}
\includegraphics[width=\columnwidth]{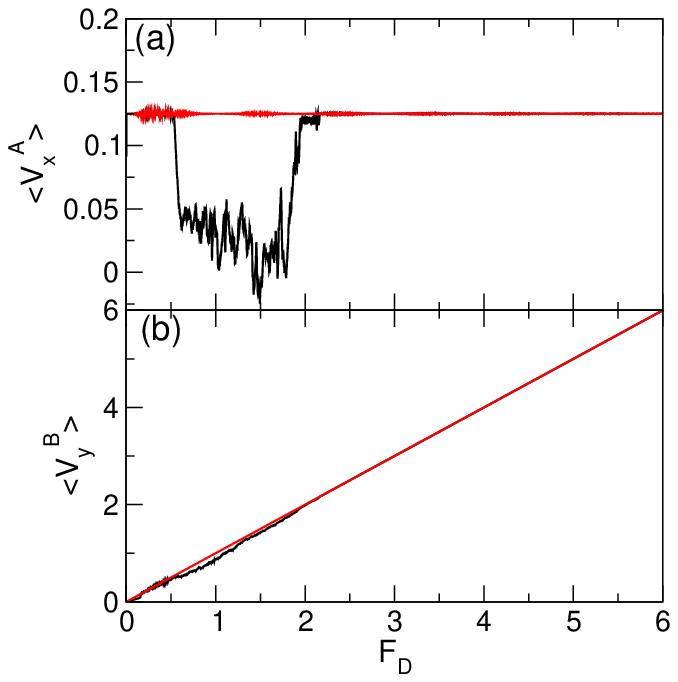}
\caption{The perpendicularly driven system with $\rho=0.44$
from Fig.~\ref{fig:22} where species A
is driven at $F_D^A=0.25$ along $+x$ and species B is driven along $+y$ at
$F_D$.
Black (red) curves are for the ramp-up (ramp-down) of $F_D$.
(a) $\langle V_x^B \rangle$ vs  $F_D$. 
(b) $\langle V_y^B \rangle$ vs  $F_D$.
}
\label{fig:23}
\end{figure}

In the previous section, we applied perpendicular driving where species A
was driven along $+x$ at fixed
$F^A_D = 1.5$ and species B was driven along $+y$ at varied $F_D$.
We next consider the effect of changing the fixed value of $F^A_D$ in the
same drive configuration.
In the limit $F^A_D = 0.0$,
the system forms a laned state in which species B
particles move between the mostly immobile species A
particles, and the lanes are aligned in the $y$-direction.

\begin{figure}
\includegraphics[width=\columnwidth]{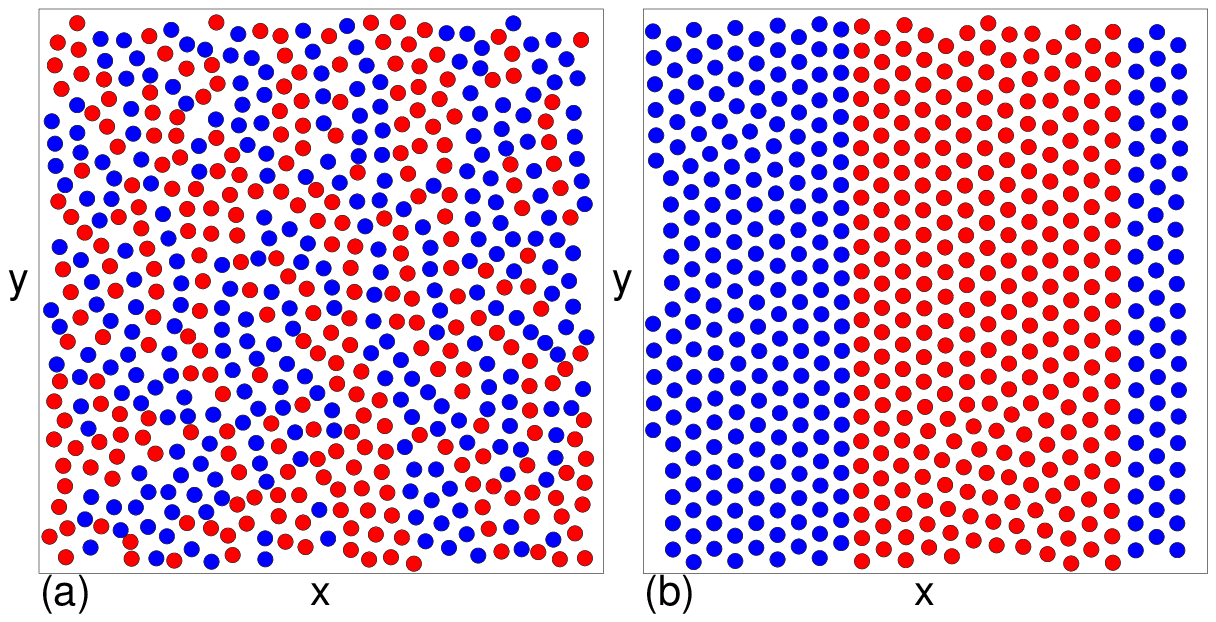}
\caption{Particle positions in
the perpendicularly driven system with $\rho=0.44$
from Fig.~\ref{fig:22} where species A (blue)
is driven at $F_D^A=0.25$ along $+x$ and species B (red) is driven along $+y$ at
$F_D$.
(a) The disordered flow phase at $F_D = 1.0$.
(b) The perpendicular-laned state at $F_D = 4.0$.
}
\label{fig:24}
\end{figure}

In Fig.~\ref{fig:22}(a),
we plot $\langle V_x^A \rangle$ versus $F_D$
for a system with $\rho = 0.44$ and $F^A_D = 0.25$ for the ramp-up
and ramp-down cycle. The corresponding
$\langle V_y^A\rangle$ and $P_6$ versus $F_D$ curves appear in
Fig.~\ref{fig:22}(b,c), respectively, while
Fig.~\ref{fig:23}(a,b) shows $\langle V_x^B \rangle$ and
$\langle V_y^B \rangle$, respectively,
versus $F_D$.
In Figs.~\ref{fig:22} and \ref{fig:23},
for $F_{D} < 0.4$ during the ramp-up,
$\langle V_x^A \rangle = 0.125$ and
$\langle V_x^B \rangle = 0.125$,
indicating that species B is being dragged by species A
along the $+x$ direction.
For $F_{D} > 0.16$, the system acts as a rigid solid,
and all of the particles move as a unit.
For $0.16 < F_{D} < 0.4$,
there is a partial decoupling of species A and B,
where species A moves only in the $+x$-direction,
while species B is moving in both the $+x$ and $+y$ directions.
For $0.4 < F_{D} < 2.2$, the system enters
a disordered flow phase, as shown in
Fig.~\ref{fig:24}(a) at $F_{D} = 1.0$.
Within the disordered flow regime,
$\langle V_y^A \rangle$ is high due to
a dragging effect from species B on species A.
We find that $\langle V^A_x \rangle$ increases
with increasing $F_D$ in the disordered phase,
while $\langle V^B_x \rangle$ decreases
and $\langle V^B_y \rangle$ increases linearly but shows fluctuations.
For $F_{D} > 2.0$, a fully phase-separated, perpendicular-laned state
appears, as shown in Fig.~\ref{fig:24}(b) at
$F_{D} = 4.0$.
Here the particles
form a mostly triangular lattice, as indicated by the fact that
$P_{6} = 0.98$ in Fig.~\ref{fig:24}(b).
Species A is moving only in the $x$-direction, giving
$\langle V^A_y \rangle = 0.0$ in Fig.~\ref{fig:22}(a),
and travels with a constant $\langle V^A_x \rangle = 0.125$.
At the same time, species B has a constant
$\langle V^B_y \rangle = 0.125$,
and the motion of species A and B is
locked along the $x$-direction.
The total velocity of species B 
increases linearly in the perpendicular-laned state with increasing
$F_{D}$.
During the ramp-down, the perpendicular-laned state persists
all the way down to $F_{D} = 0.0$,
and on the next ramp-up, the system does not show any hysteresis.

\begin{figure}
\includegraphics[width=\columnwidth]{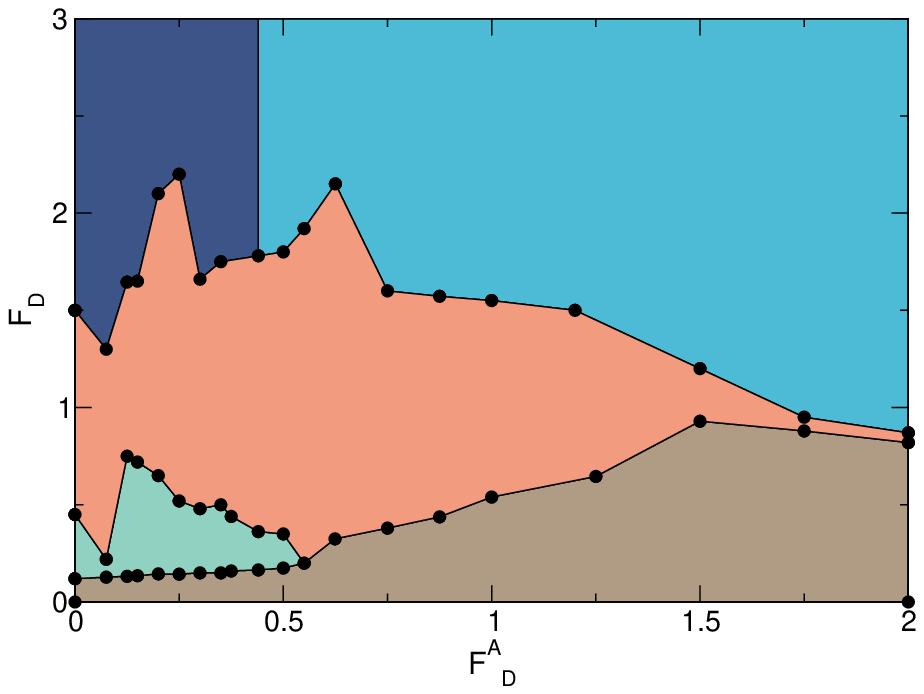}
\caption{Dynamic phase diagram as a function of
$F_D$ vs $F_D^A$ for the perpendicularly driven system from
Fig.~\ref{fig:22} with $\rho = 0.44$.
The phases are: locked (brown),
1D flow (green),
disordered flow (orange),
perpendicular laned state (dark blue),
and the tilted lanes state on the ramp-up portion of the drive
(light blue).}
\label{fig:25}
\end{figure}

\begin{figure}
\includegraphics[width=\columnwidth]{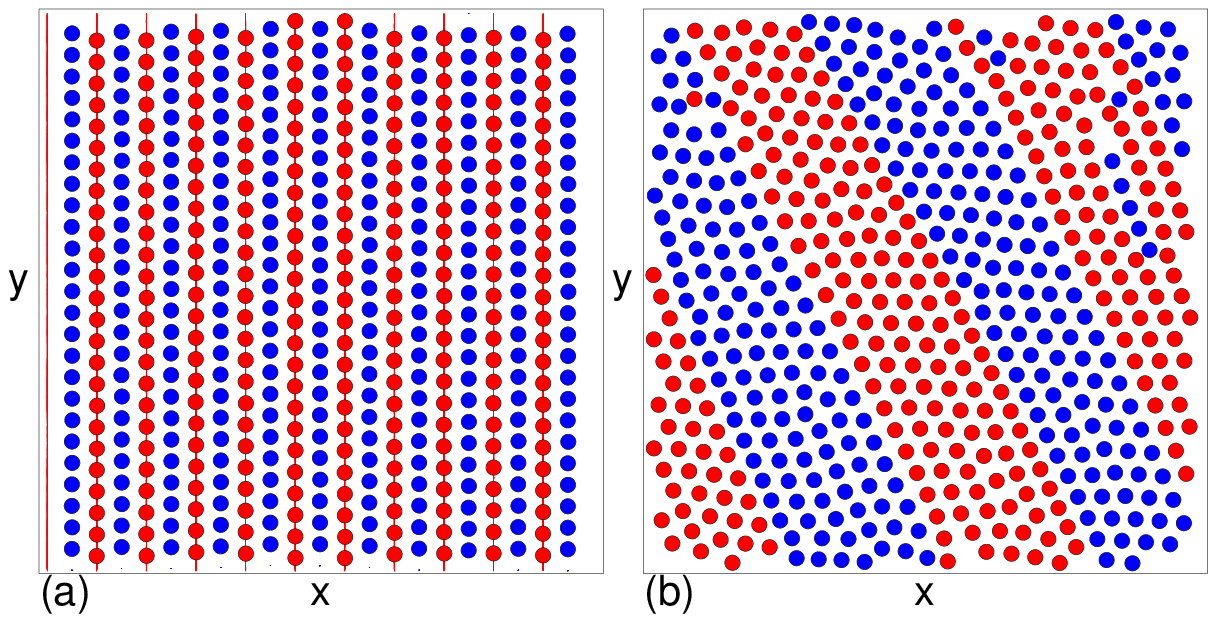}
\caption{(a) Particle positions and trajectories
in the perpendicularly driven system with $\rho=0.44$ from
Fig.~\ref{fig:25} at $F^A_D = 0.0$ and $F_D=0.0$
where lanes of mobile species B particles (red) are separated by
lanes of stationary species A particles (blue).
(b) Particle positions for the same system
in the laned state at $F_D = 2.0$ and $F^A_D = 0.5$.
}
\label{fig:26}
\end{figure}

In Fig.~\ref{fig:25}, we plot a dynamic phase diagram
as a function of $F_D$ versus $F^A_D$ at
$\rho = 0.44$ for the perpendicularly driven
system from Fig.~\ref{fig:22}.
We highlight the locked phase, the 1D motion phase,
the disordered flow phase, the perpendicular laned state,
and the tilted lanes state that appears during
the ramp-up portion of the drive.
For $F^A_D > 0.5$, the 1D motion is lost,
and for $F^A_D > 0.4$, the perpendicular laned state vanishes.
At $F^A_D = 0.0$, only species B particles
are driven, and we find
a 1D laned state with mobile lanes of
species B flowing along the y-direction
separated by stationary lanes of species A,
as shown in Fig.~\ref{fig:26}(a) at $F^A_D = 0.0$.
As $F_D$ increases and
the system enters the disordered flow state,
the motion shows strong fluctuations and is
partially aligned with the drive, but not fully aligned
with the drive as in the laned states.
At large $F_D$, phase segregation into the perpendicular laned
state illustrated in Fig.~\ref{fig:24}(b) occurs.
For $F^A_D > 0.5$, a tilted lane state appears instead,
as shown in Fig.~\ref{fig:26}(b) at $F^A_D = 0.5$ and $F_D=2.0$.

\begin{figure}
\includegraphics[width=\columnwidth]{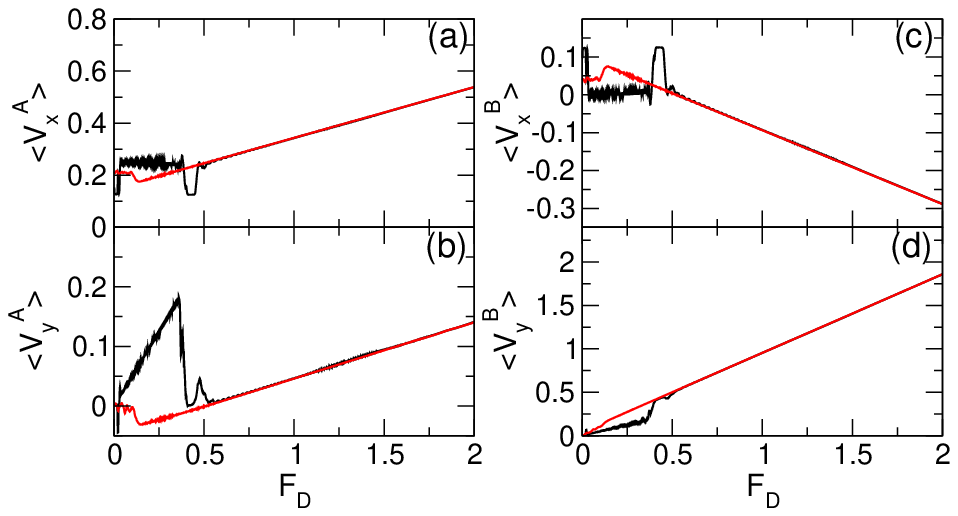}
\caption{(a) $\langle V^A_x \rangle$ and (b)
$\langle V^A_y\rangle$ vs $F_{D}$
for a perpendicularly driven system with $\rho = 0.0926$ and
$F^A_D=0.25$.
Black (red) curves are for the ramp-up (ramp-down) of $F_D$.
(c,d) The corresponding (c) $\langle V^B_x \rangle$ and
(d) $\langle V^B_y \rangle$ vs $F_{D}$.
}
\label{fig:27}
\end{figure}

\begin{figure}
\includegraphics[width=\columnwidth]{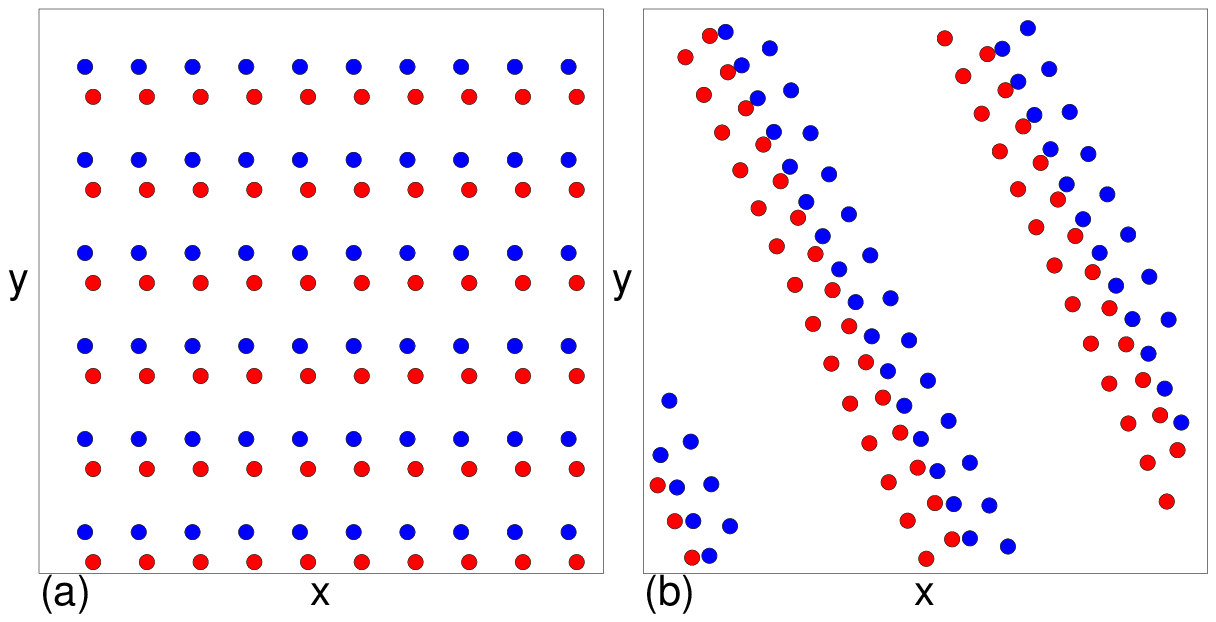}
\caption{Particle positions in the perpendicularly driven system
with $\rho=0.0926$ from Fig.~\ref{fig:27} where species A (blue) is
driven at $F^A_D=0.25$ along $+x$ and species B (red) is driven
along $+y$ at $F_D$.
(a) The locked state at $F_{D} = 0.3$, where stripe-like structures
aligned in the $x$-direction appear.
(b) A tilted lane state at $F_{D} = 2.0$.
}
\label{fig:28}
\end{figure}

The perpendicular laned state persists for higher
particle densities $\rho$ but does not occur at lower $\rho$.
In Fig.~\ref{fig:27}(a,b), we plot
$\langle V^A_x \rangle$ and $\langle V^A_y\rangle$, respectively,
versus $F_D$ for a perpendicularly driven
system with $\rho = 0.0926$ and fixed $F^A_D = 0.25$.
Figure~\ref{fig:27}(c,d) shows the corresponding
$\langle V^B_x \rangle$ and $\langle V^B_y \rangle$,
respectively, versus $F_D$.
For drives below $F_D=0.35$, there are no laned or locked states;
instead, we find
a partial 1D flow state followed by a tilted laned state.
At lower particle densities such as this one, the locked state
becomes increasingly stripe-like in character,
as shown in Fig.~\ref{fig:28}(a) at $F_D = 0.3$, while
a laned state emerges at higher drives, as
illustrated in Fig.~\ref{fig:28}(b) at $F_D = 2.0$.
There is still strong hysteresis across the laned states
during the ramp-down.

\section{Discussion}

In this work we considered the case of 50:50 mixtures of species A and B,
but we expect other effects could arise if this ratio were varied.
For example, a minority species could remain
locked in the driving direction of the majority species at high
drives, or laning could be suppressed.
We have focused on dc driving,
but it would be interesting to explore
what would happen for ac driving, and how this might alter
the formation and/or appearance of the tilted lane structures.
It would also be possible to introduce
mixtures of three species moving in three different directions,
in which case the system could remain disordered for larger drives,
or could form more complex laning structures.
We did not include thermal fluctuations, but we expect that in the presence
of a finite temperature, similar phases would appear but the
the disordered regions could be expanded, while at higher temperatures,
the systems could become completely disordered.
There could still be laning at higher temperatures,
but the triangular ordering that we observe
in the individual lanes could be lost.

\section{Summary}

We have examined hysteresis and laning transitions in
binary mixtures of repulsively interacting particles driven in
different directions. When species A and species B are driven in
opposite directions, we find a jammed state,
a disordered flow state, and a laned state as a function of
increasing drive.
There is pronounced hysteresis across the laned state between the
ramp-up and ramp-down of the drive, with
the laned state persisting down to much lower drives
during the ramp-down.
This hysteresis remains robust for a range of particle densities.
When species B is driven perpendicularly to species A,
we find a low-drive locked phase where the two species move together at
45$^\circ$, a decoupled 1D flow phase, a disordered flow phase,
and a laned state with tilted lanes.
There is strong hysteresis in the tilted laned state between the
ramp-up and ramp-down of the drive.
We have also considered a
perpendicular driving protocol in which species A experiences a
fixed $x$-direction drive while a $y$-direction drive on species B is gradually
increased. This leads to the formation of
a series of different laned states in which the lanes are roughly oriented
in the direction of the net external drive.
The tilted laned states can produce a negative drag effect in which species
A particles
move in the direction opposite to the
driving direction of the species B particles
due to a deflection of the species A particles by species B lanes.
Once a tilted lane state forms, it can become compressed into a
stripe state, where there is a discernible transition into a new,
higher-angle tilted state as the $y$-direction drive on species B increases.
During the ramp-down, these high-angle tilted lane states
persist to significantly lower drives, and are destroyed in
disordering transitions.
When the fixed drive on species A is small,
a perpendicular laned state can appear on the ramp-up of the
$y$-direction drive on species B.
We show that the transitions between the ordered and disordered flows and
and among the different laned states are associated with
jumps in the different components of the velocity-force curves and in
the fraction of topological defects, as well as with
strong hysteresis.

\begin{acknowledgements}
We gratefully acknowledge the support of the U.S. Department of
Energy through the LANL/LDRD program for this work.
This work was supported by the US Department of Energy through
the Los Alamos National Laboratory.  Los Alamos National Laboratory is
operated by Triad National Security, LLC, for the National Nuclear Security
Administration of the U. S. Department of Energy (Contract No. 892333218NCA000001).
\end{acknowledgements}

\bibliography{mybib}

\end{document}